\documentclass[prd,twocolumn,amsmath,amssymb]{revtex4-2}
\usepackage{graphicx}
\usepackage{amssymb}
\usepackage{amsmath}
\usepackage[braket, qm]{qcircuit}
\usepackage{float}
\usepackage{bbold}
\usepackage{appendix}
\usepackage{comment}

\newcommand{\nmax}{n}
\DeclareMathOperator{\Trace}{Tr}
\usepackage{bbm}
\usepackage{lineno}
\usepackage{slashbox}
\usepackage{xcolor}
\begin{document}
\title{Multipartite entanglement for d-level systems in 1+1D}

\author{Zane Ozzello}
\author{Yannick Meurice}
\affiliation{Department of Physics and Astronomy, The University of Iowa, Iowa City, IA 52242, USA }

\date{\today}

\begin{abstract}

We show that multipartite entanglement can be used as an efficient way to identify the critical points of 1+1D systems.  We demonstrate this with the quantum Ising model, lattice $\lambda \phi^4$ approximated with qutrits, and arrays of Rydberg atoms.  To do so, we use multipartite compositions of entanglement quantities -- $S_A$,$S_B$,$S_C$,$S_{AB}$,$S_{BC}$,and $S_{ABC}$ -- combined to form the strong subadditivity, weak monotonicity, and a convex combination of these with conformal properties.  We will demonstrate how these bipartite entanglement quantities of a quadpartite system display behavior at phase boundaries, but the combination of these aforementioned quantities sharpens and localizes this behavior at the boundaries even better.  We will show that we can extend a scheme for approximating the entanglement with the mutual information, and that this acts as a lower bound for the individual bipartite quantities.  We show the mutual information will also follow the changes in the entanglement for the above combined quantities, despite no longer being a strict lower bound and the additional contributions of different signs.  This mutual information approximation to the identifying quantities can have its lower probabilities removed in a process we call filtering, and despite the combination of terms with different signs, will respond well to the filtering and offer improvements to the lower bound.  
 
 \end{abstract}

\maketitle

\def\beq{\begin{equation}}
\def\enq{\end{equation}}
\def\nq{n_q}
\def\nmax{n_{\mathrm{max}}}
\def\phix{\hat{\phi} _{\bf x}}
\def\nq{n_q}
\def\ah{\hat{a}}
\def\ahd{\hat{a}^\dagger}
\def\pn{P_{\nmax-1}}
\def\har{\hat{H}^{\mathrm{har.}}}
\def\hanh{\hat{H}^{\mathrm{anh.}}}
\def\hpf{\hat{\phi}^4}
\def\np{\mathcal{N}_p}

\section{Introduction}
In recent years, there has been strong interest regarding entanglement entropy within gauge theories \cite{ent_gauge,ent_distill_gauge,confine_tensor,ent_dp1gt,ent_w_elecstring,ent_lat_decomp,ssa_in_lgt,su3_4d_gt,maxwell_sphere,ee_confine_workaround}.  This is partially motivated by an interest in calculating the real-time evolution of lattice gauge theories \cite{dis_entprobe,dis_entprod,eff_basis_entscale,Schuhmacher:2025ehh,Surace:2020ycc,Davoudi:2025rdv,Tan:2019kya,Verdel:2019chj}.  Such digitized or truncated models are designed with implementations on quantum devices or analog simulators in mind.  Some of these have been known to have a very rich phase diagram in their space of tunable parameters \cite{Banerjee:2015pnt,Keesling:2018ish,Zhang:2023agx,rydphase_nature,NSSrivatsa:2025jhh}.  Entanglement can be a tool for understanding such phase diagrams.  However, entanglement is pervasive within current research beyond involvement in lattice gauge theories in this way; we see applications utilizing both the bipartite entanglement entropy \cite{bip_review} concerning the entanglement of one part of the system with another and the multipartite entanglement entropy \cite{multip_review} concerning the entanglement of multiple parts of the system with one another.  Entanglement plays a role in other areas of physics such as scattering \cite{dis_entprobe,dis_entprod,ent_pp_coll}, many-body physics \cite{mbl_ent,mbs_entangle}, nuclear physics \cite{nuclear_bases,Brokemeier:2024lhq}, conformal field theories (CFTs) \cite{sdelta, sdelt2}, and many other areas of physics \cite{arealaw_rev,mps_ent,maloney_monogamy, mb_crit_spin, mp_edge_and_tensor, mp_eth, mp_q_anomaly, mp_holo_and_ineq, mp_holo_pure,pspace_ent,multipart_phase_wfisher_hauke_zoller}.  

Entanglement entropy, since it involves the interconnectedness of a system, can be very difficult to experimentally measure, causing problems not just in the multipartite case, but also in the bipartite case \cite{lukin_enttest, multinode_multipart}.  In the age of quantum computers, a work around to these limitations is of premier interest.  Potentially, by having a means of identifying entanglement with a quantum computer, there could be a measurement of the quantum properties of a quantum system directly.  As a result, finding means to approximate the entanglement while its direct measurement is difficult is paramount \cite{multipart_phase_wfisher_hauke_zoller,multitop_founda}.  Recently, methods have been proposed to utilize classical quantities, Shannon entropy and mutual information, to approximate bipartite entanglement entropy \cite{yannick_bitstring}.  This is done by using bitstring results from a quantum computer, along with their associated counts divided by shot totals, to create probabilities for the aforementioned quantities.  In this work, we take the methods a step further, generalizing and showing that the methodology is applicable to multipartite entanglement entropy.  Despite the fact that this filtering is applied to the multiple mutual information quantities at once, which introduce even more Shannon entropies at once, the filtering still yields plateaus that could be used as potential improvements to the bound.  

In this article, we apply entanglement entropy in the areas of conformal field theory and phase transitions \cite{kitaev_crit_phenom, cardy_cft, multipart_phase_wfisher_hauke_zoller}.  We will further show the efficacy of our ditstring approach to multipartite entanglement, we take the quantities proven to be greater than zero in the inequalities of strong subadditivity and weak monotonicity, then take a convex combination of these with conformal properties, $S_\Delta$ which is outlined in \cite{sdelta} in the context of conformal field theory.  We will use these quantities to identify critical points using entanglement entropy and its approximation.  Motivated by the very clear correspondence between the phase diagrams for a Rydberg simulator shown in Figure 1.B, made with structure factors in the main text of Ref. \cite{rydphase_nature}  and the entanglement entropy in Figure 6 of the supplemental materials of the same article, we are also able to empirically identify the critical points of different models with our method. However, we are not aware of rigorous results supporting this observation.

These quantities are guaranteed to be positive and are a set of entanglement entropies of different parts with each other that are added and subtracted to form the quantities.  Despite these combinations of positive and negative terms, these quantities manage to isolate a peak in a parameter range by canceling off values to either side.  Not only does the peak occur in this way, but such cancellations also better isolate these quantities to define the phase boundaries for their respective models.  We will further highlight these behaviors in Sec. \ref{sec:multip}.  All results in this work are found using exact diagonalization, but all methods surrounding probabilities used in the mutual information can be applied to ditstring probability returns from quantum devices.  

The article is organized as follows.  In Sec. \ref{sec:hilbert} we discuss the structure of multipartite Hilbert space and its application to von Neumann entanglement entropy and mutual information.  In Sec. \ref{sec:multip}, we discuss how we attempt to find critical points with the ideas of \cite{sdelta} as our starting point.  We then apply our multipartite scheme and ditstring approximation to three 1+1D models:  quantum Ising, lattice $\phi^4$ with qutrits, and a chain of Rydberg atoms.  For this, we work exclusively in periodic boundary conditions.  When we loosen this restriction, we look at the impact of switching to open boundary conditions in Sec. \ref{sec:obc}.  We then utilize the methods laid out in \cite{filter_bits, exploring} in Sec. \ref{sec:filt} and apply a filtering scheme that involves the removal of low probabilities that go into the composite entropic quantities.  In the conclusions, we touch on future directions, looking at the validity of this method for identifying critical points, further pushing the mutual information lower bound, and discussion as to how these can be combined for actual quantum computer implementation.  In the appendices, we discuss how the choice of basis impacts the value of the mutual information, provide added details surrounding the conditions of efficacy for the critical point finding method, and discuss a different implementation of the chain of Rydberg atoms.

\section{Multipartite Qudit Hilbert Space}

\label{sec:hilbert}

In the following, we consider $N$ qudits with $d$ states in the local Hilbert space.
The computational basis then consists of $d^{N}$ elements of the form
\beq
\ket{\{n\}}\equiv\ket{n_0, n_1,\dots,n_{N_q-1}},
\enq
with $n_j=0,\dots,d-1$. Any element of this basis can be factored in a multipartite way
\beq
\label{eq:multip_n}
\ket{\{n\}}=\ket{\{n\}_A}\ket{\{n\}_B}\ket{\{n\}_C}\dots,
\enq
with
\begin{eqnarray}
\ket{\{n\}_A}&\equiv&\ket{n_0, n_1,\dots,n_{|A|-1}} \ {\rm and } \notag \\
\ket{\{n\}_B}&\equiv&\ket{n_{|A|},\dots,n_{|A|+|B|-1}} \\
\dots \notag
\end{eqnarray}
with $|A|$, $|B| \dots$ being the number of qudits in parts $A, \ B,\dots$.
An arbitrary state in the computational basis is of the form
\beq
\ket{\psi}=\sum_{\{n\}} c_{\{n\}}\ket{\{n\}}, 
\enq
and the state $\ket{\{n\}}$ will be observed with a probability
\beq
p_{\{n\}}=|c_{\{n\}}|^2.
\enq

We adopt the notation that a given system can be split into two regions, $R$ and $\bar{R}$, a set of sites and its complement.  A density matrix for the whole system is written as 
\begin{equation}
\rho_{R\bar{R}}=\ket{\psi}\bra{\psi}.
\end{equation}
So, when diagonal, becomes
\begin{equation}
\rho_{R\bar{R},diag}=\sum_{\{n\}}|c_{\{n\}}|^2 \ket{\{n\}}\bra{\{n\}}.    
\end{equation}

Our focus will be on using a pure state for a density matrix by which we can investigate the entanglement entropy of a partitioned system.  To deal with a partition, the system is split using the basis factoring laid out in Eq. (\ref{eq:multip_n}) for a two-parted system.  If we want the entanglement between two such parts, we will need to define the bipartite entanglement entropy.  To find the entanglement between two parts of a system, a reduced density matrix is needed, 
\beq
    \label{eq:reddensity_mat}
    \rho_R= \Trace_{\bar{R}} \rho_{R\bar{R}},
\enq
which can then be used to define the bipartite von Neumann entanglement entropy as
\begin{equation}
    \label{eq:vnent_red}
    S^{vN}_R \equiv -\Trace (\rho_{R}\ln{(\rho_{R})})= -\sum_{r} \lambda_{r} \ln{\lambda_{r}}
\end{equation}
where $\lambda_{r}$ are the eigenvalues of the reduced density matrix.  If we were to take $R$ as the whole system then $S^{vN}_R=0$, but we will only be interested in systems such that there are distinct $R$ and $\bar{R}$.  

The form in Eq. (\ref{eq:vnent_red}) should be reminiscent of the Shannon entropy,
\beq
\label{eq:whole_shan}
H_{R\bar{R}}\equiv -\sum_{\{n\}} p_{\{n\}}\ln(p_{\{n\}}).
\enq
Where Eq. (\ref{eq:whole_shan}) is associated with the state $\ket{\psi}$, and the probabilities involved can be estimated by ditstring returns from a quantum computer by taking the counts of each state's occurrence divided by the total shots on the system.    However, the bipartite entanglement utilizes a reduced density matrix for the entanglement of one part with another.  A reduced probability, commonly known as a marginal probability, which in a bipartition sums over the $\bar{R}$ parts while fixing the $R$, can be defined as
\begin{equation}
    \label{eq:redprob}
    p_{\{n\}_R} = \sum_{\{n\}_{\bar{R}}} p_{\{n\}_R \{n\}_{\bar{R}}},
\end{equation}
using these probabilities with the Shannon entropy (Eq. (\ref{eq:whole_shan}))
would then lead to what will be referred to as a reduced Shannon entropy for one part of the whole:  
\begin{equation}
    \label{eq:shan_onepart}
    H_R \equiv -\sum_{\{n\}_R} p_{\{n\}_R} \ln{p_{\{n\}_R}}.  
\end{equation}
This reduced Shannon entropy will then be used in the quantity to approximate the von Neumann entanglement entropy, the mutual information of the ``experimental'' probabilities:  
\begin{equation}
    \label{eq:mi}
    I_{R\bar{R}} \equiv H_R + H_{\bar{R}} - H_{R\bar{R}}.
\end{equation}
The mutual information informs us how much would be learned about $\bar{R}$ by learning what $R$ is, or vice versa \cite{witten_inforev}, which naturally falls into line with the thinking of entanglement entropy as a way of quantifying the quantum relationship between partitionings of a system.  At this point, it should be clear that based on how the bipartite entanglement is defined, the entanglement of a region with its complement is the same as the entanglement of a complement with the region, meaning $S_R=S_{\bar{R}}$.  However, as evident from its definition, the reduced Shannon entropy does not guarantee the same, meaning $H_R \neq H_{\bar{R}}$.  

The multipartite case is simply an extension of the bipartite case, which we have handled generally until now.  One can treat a system multipartitioned as a combination of bipartitions.  This follows naturally from the idea that $S_R=S_{\bar{R}}$.  So, by finding a handful of individual bipartite entropies for a system, we can implicitly find those of the complements.  We show examples of this up to the case we primarily consider in this work, a multipartite system in four parts, in Tab. \ref{tab:multip_parts}.  
\begin{table}
\caption{Bipartite entropies given certain amounts of parts for a  system \label{tab:multip_parts}}
\begin{tabular}{ |p{2.5cm}||p{2.5cm}|p{2.5cm}|  }

 \hline
 \multicolumn{3}{|c|}{Multipartite Partitioning} \\
 \hline
 Parts& Explicitly Chosen &Implicitly Found\\
 \hline
 A,B  & $S_A$    &$S_B$\\
 \hline
 A,B,C&$S_A$, $S_B$, $S_C$  & $S_{BC}$, $S_{AC}$, $S_{AB}$\\
 \hline
 A,B,C,D & $S_A$, $S_B$, $S_C$, $S_D$, $S_{AB}$, $S_{AC}$, $S_{BC}$& $S_{BCD}$, $S_{ACD}$, $S_{ABD}$, $S_{ABC}$, $S_{CD}$, $S_{BD}$, $S_{AD}$\\
 
 \hline
\end{tabular}
\end{table}
According to the Holevo bound, each one of these von Neumann bipartite entanglement entropies are bounded below by their mutual information equivalents.  However, if we were to add or subtract any entanglement entropic quantities, there is no guarantee that their mutual information analogs are bounded in the same way.

Before progressing to the next section, we will establish a few notational choices that have been made for convenience throughout the work.  First, the von Neumann entanglement entropy of a partition $R$ with the rest of the system will be denoted as $S_R$.  For the mutual information, we will treat the second partition as implicit; e.g. for the mutual information between $A$ and $B$, we will use $I_{R\bar{R}}=I_{R}=I_{\bar{R}}$, a notational form matching the entanglement entropy convention.  Additionally, in the next section we will be introducing entropic quantities which are composed of multiple bipartitions.  Given that this would require each term to have labeling for all four parts, we will use the implicit notation for the mutual information to prevent equations from being too busy.

\section{Multipartitions in Different Models}
\label{sec:multip}

Beyond just looking at bipartitions of a system, we want to further test our multipartite capabilities with implementations of composite quantities.  We begin within the context of the Entanglement Bootstrap approach of conformal field theory \cite{kim_boostrap_4delt}.  Lin and McGreevy \cite{sdelta}, 
showed that a special linear combination of entanglement Hamiltonians of the form $K_X=-\log(\rho _X)$ has remarkable properties. The reduced density matrices $\rho _X$ are obtained by tracing over the complement of
the subsets $A,\ B,\ C,\ AB,\ BC$ and $ABC$ of a general multipartiton $ABCD$.  From these $K_X$ they build a special linear combination, 

\begin{align}
\label{eq:kdelt}
K_\Delta &:= (K_{AB}+K_{BC})- \eta (K_A+K_C) \notag \\
&-(1-\eta)(K_B+K_{ABC})    \end{align}
which can be satisfied by a ground state of a 1+1D unitary CFT.  In addition, the authors showed that the corresponding linear combination of entanglement entropies to the previous,
\begin{align}
    \label{eq:sdelt}
    S_{\Delta} &:= (S_{AB}+S_{BC})- \eta (S_A+S_C) \notag \\ &-(1-\eta)(S_B+S_{ABC})  
\end{align}
where
\begin{equation}
    \label{eq:eta}
    \eta = \frac{(x_2-x_1)(x_4-x_3)}{(x_3-x_1)(x_4-x_2)}
\end{equation}
with partition edges $x_i$, has fascinating extremal properties for such a CFT ground state as mentioned above.  Furthermore, $S_\Delta$ is by definition positive as a convex combination of the nonnegative quantities $S_{weak}$ and $S_{strong}$ from information theory \cite{nandc}.  These quantities are defined and approximated with the mutual information as, respectively,
\begin{equation}
    \begin{split}
    \label{eq:weak}
        S_{weak} &= S_{AB} + S_{BC} - S_A -S_C \geq 0\\
        &\simeq I_{AB}+I_{BC}-I_{A}-I_{C}\\
        &=H_{AB}+H_{CD}+H_{BC}+H_{AD}\\
        &-H_{A}-H_{BCD}-H_C-H_{ABD} 
    \end{split}  
\end{equation}
and
\begin{equation}
\begin{split}
    \label{eq:strong}
    S_{strong} &= S_{AB} + S_{BC} - S_B - S_{ABC} \geq 0\\
        &\simeq I_{AB}+I_{BC}-I_{B}-I_{ABC}\\
        &=H_{AB}+H_{CD}+H_{BC}+H_{AD}\\
        &-H_{B}-H_{ACD}-H_{ABC}-H_{D}.
    \end{split}
\end{equation}
Evidently, $\eta$ interpolates $S_{\Delta}$ between strong subadditivity at $\eta=0$ and weak monotonicity at $\eta=1$.  

Through observation, we have found that when varying over a critical parameter and calculating the $S_{\Delta}$, weak monotonicity, or strong subadditivity quantity at each step, these quantities maximize at a parameter value that coincides with the occurrence of a critical point for the system.  We will show evidence for this behavior with the forthcoming models, both with the von Neumann entanglement entropy and the lower-bound approximation mutual information.  Note that for full efficacy of this scheme, we will require periodic boundary conditions and the system to be split into consecutive parts.  The impact of not following these restrictions will be touched on in Sec. \ref{sec:obc}.  In App. \ref{sec:app_sdelt_dep}, we show how other choices of system size and subsystem composition impact the critical point identification with either open or periodic boundary conditions.

There are also cases where $S_{weak}$ (Eq. \ref{eq:weak}) may be equal to $S_{strong}$ (Eq. \ref{eq:strong}).  This is a result of how the constituent bipartite quantities combine comparatively in the composite quantities.  For example, a partition of AABBBCCCDD would cause $S_{weak}=S_{strong}$.  Both quantities have $S_{AB}$ and $S_{CD}$, so they are naturally the same.  Also, $|B|=|C|$ so therefore $S_B=S_C$, but $|A|=2$ and $|ABC|=8$.  However, the entanglement between a region and its complement should be equivalent, so given $|A|$ and $|ABC|$ are the size of each other's complement, $S_{ABC}=S_A$ and thus $S_{weak}=S_{strong}$ for this particular partitioning.  There are also cases where $S_{weak}$ is exactly proportionate to $S_{strong}$.  Using a similar argument as above, we can show that $S_{weak}=2S_{strong}$ for a partition DDAABBCCDD.  

It should be noted that we will see some slight differences in behavior of the mutual information for these composite quantities, as opposed to singular quantities such as $S_R$.  The Holevo bound \cite{holevo} indicates that the mutual information should strictly be a lower bound to the entanglement \cite{nandc,preskill_qi}.  Looking at the quantity $S_{strong}$, this guarantees that $I_{AB}<S_{AB}$, $I_{BC}<S_{BC}$, $I_{B}<S_{B}$, and $I_{ABC}<S_{ABC}$ are always true, but it does not make such guarantees about how the mutual information quantities will behave with respect to the entanglement entropy when combined together as a result of the differing signs on the quantities involved.  Therefore, if data shows the mutual information being greater than the entanglement, this does not mean the Holevo bound is violated, as the bound makes no guarantees about how combinations should behave.  This follows directly from the alternating signs in the definitions of our composite multipartite quantities, as if they were all positive, we would still be able to treat the mutual information approximation to the entanglement as a lower bound.  

\subsection{1D Quantum Ising}
\label{subsec:ising}
The transverse-field Ising model in one spatial dimension has two common representations \cite{ising_scatter}.  The first being in the particle, or Z, basis:
\beq
    \label{eq:ising_part}
        \hat{H}_\text{particle} = - h \sum_{i=1}^{N_s} \hat{\sigma}^z_i -J \sum_{i=1}^{N_s} \hat{\sigma}^x_{i} \hat{\sigma}^x_{i+1}.
\enq
The second in the spin, or X, basis:
\beq
    \label{eq:ising_spin}
    \hat{H}_\text{spin} = - h \sum_{i=1}^{N_s} \hat{\sigma}^x_i -J \sum_{i=1}^{N_s} \hat{\sigma}^z_{i} \hat{\sigma}^z_{i+1}.  
\enq
For both representations, $h$ is the transverse magnetic field (also commonly written as $h_T$), also known as the on-site energy.  $J$ is the ferromagnetic nearest-neighbor coupling, from this interaction comes the creation, annihilation, and particle hopping.  Both of the previous models are related by a Hadamard transformation.  We will discuss effects of these different bases on the mutual information in Appendix \ref{sec:app_mutinf}.  

This model is well studied, including work done with quantum devices.  There is a well-known second-order phase transition in the infinite-volume limit.  Here, at $J/h=1$, the model changes from a disordered phase to an ordered phase.  Typically, the Z basis is better suited for modeling in the disordered phase with its local part diagonal, and the X basis is better suited for the ordered phase with its interacting part diagonal.  Unless otherwise detailed, calculations were carried out in the Z basis.

\begin{figure}[ht!]
\centering
\includegraphics[width=4.25cm]{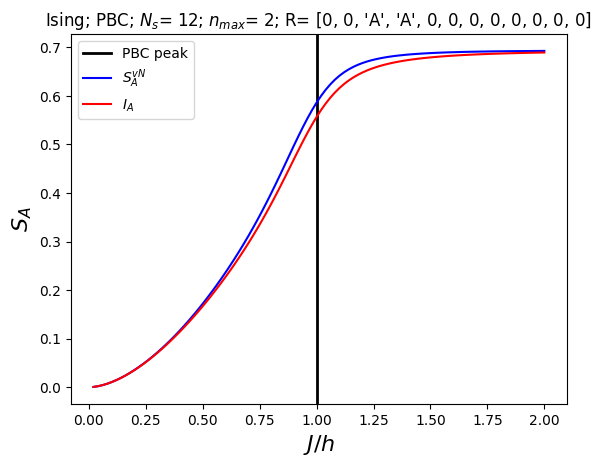} 
\includegraphics[width=4.25cm]{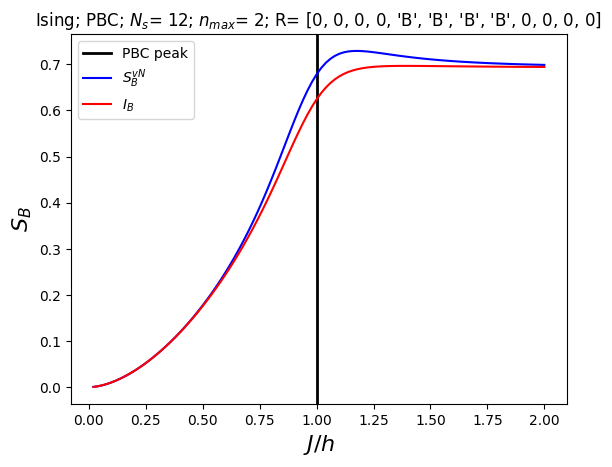}
\includegraphics[width=4.25cm]{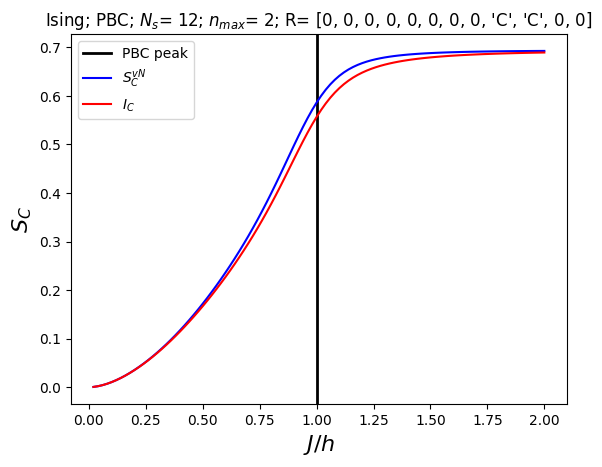}
\includegraphics[width=4.25cm]{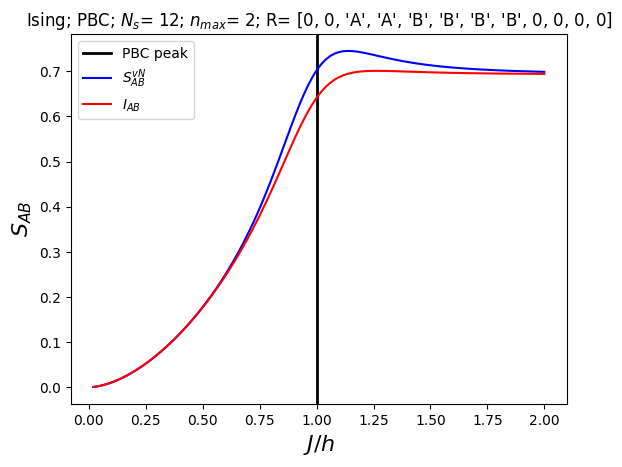}
\includegraphics[width=4.25cm]{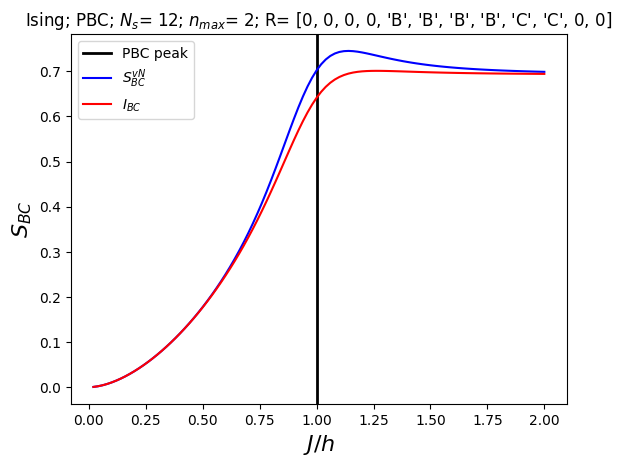}
\includegraphics[width=4.25cm]{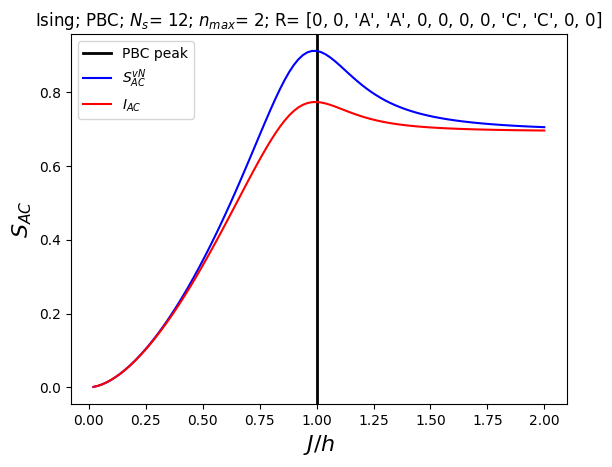}
\includegraphics[width=4.25cm]{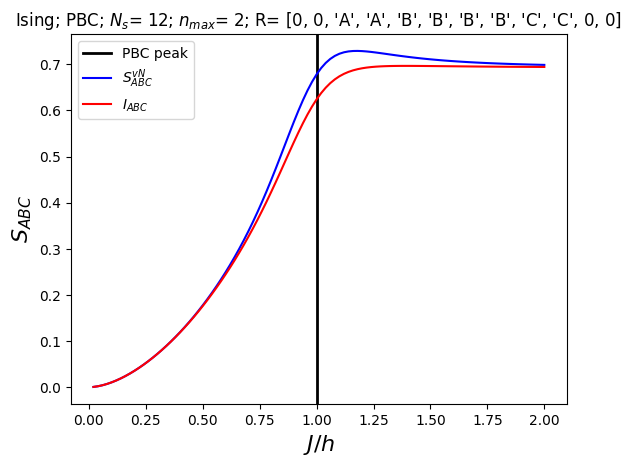} 
\caption{\label{fig:ising_parts}12-site Ising chain varying over $J/h$ showing multipartite entropy with partitioning DDAABBBBCCDD.  Critical point is marked with vertical black line.  From top left to bottom: $S_A$, $S_B$, $S_C$, $S_{AB}$, $S_{BC}$, $S_{AC}$, $S_{ABC}$.}
\end{figure}

In Fig. \ref{fig:ising_parts}, the mutual information is compared with the von Neumann entanglement entropy and empirically forms a tight lower bound.  On the plots, the known point of the infinite-volume critical point is marked with a black line for reference.  Observationally, the points where the tightness of the mutual information as a lower bound loosens are near this critical point, especially for larger partitions or separated parts.  

The lower bound of the mutual information also approximates the entanglement entropy well for the composite values of the previous inequalities, Eqs. (\ref{eq:weak}) \& (\ref{eq:strong}).  Now, some overshooting appears away from the peak, but this just arises from there being combinations of additions and subtractions of quantities.  Most importantly, the mutual information and entanglement have the same graphical behavior and both are clearly positive for all values.  It is also observed that $S_{weak}$ and $S_{strong}$ seem to have a multiplicative factor that would relate them.  

\begin{figure}[ht!]
\centering
\includegraphics[width=4.25cm]{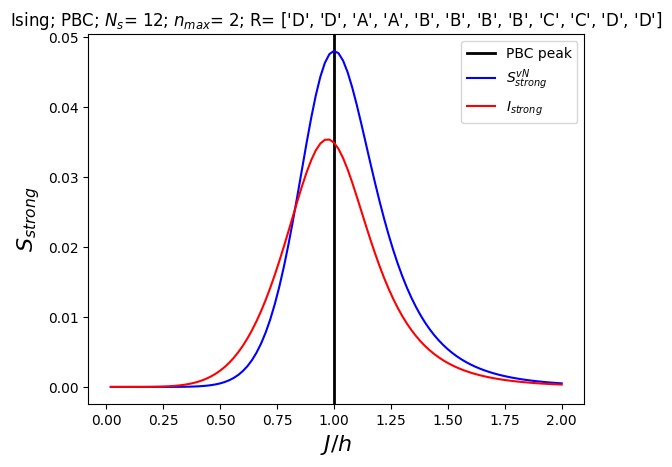} 
\includegraphics[width=4.25cm]{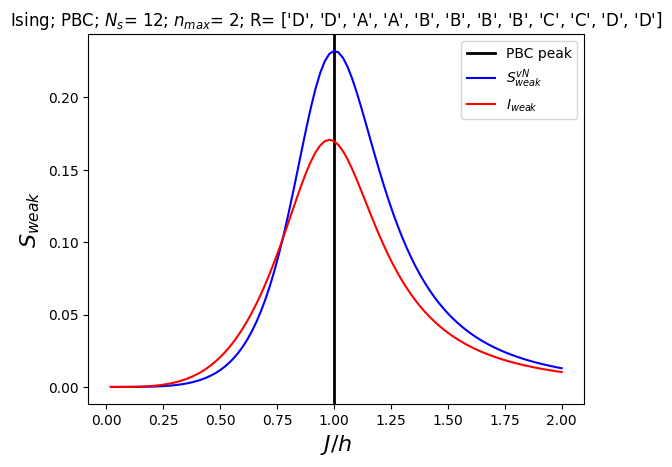}

\caption{\label{fig:ising_ineqs}12-site Ising chain varying over $J/h$ showing, from left to right, strong subadditivity and weak monotonicity with partitioning DDAABBBBCCDD.}
\end{figure}

\begin{figure}[h!]
\centering
\includegraphics[width=7cm]{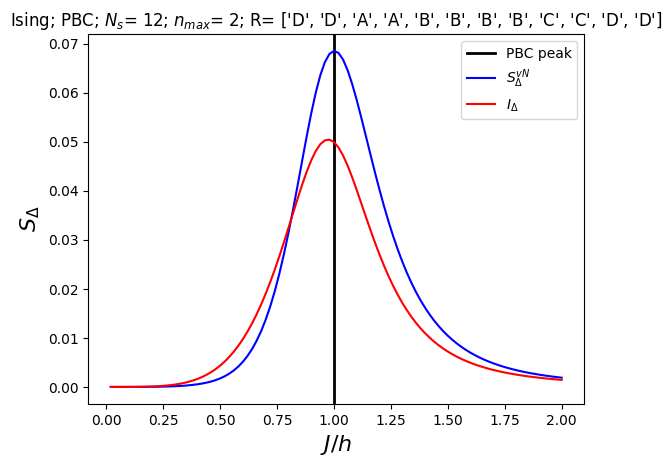}
\caption{\label{fig:ising_delta}
$S_{\Delta}$ for a 12-site Ising chain varying over $J/h$ with partitioning DDAABBBBCCDD.}
\end{figure}

These quantities obeying their respective inequalities are then combined for the $S_{\Delta}$ for the Ising case, Fig. \ref{fig:ising_delta}.  

This quantity again experiences overshooting for the same reasons as before, but again we get the results we want.  The mutual information captures the same behavior as the entanglement and is clearly positive. Importantly, the entanglement peaks at $J/h=1$, the infinite-volume critical point.  The mutual information's peak is slightly shifted towards the disordered phase.  

The way in which $S_{strong}, \text{ } S_{weak}, $ and $S_\Delta$ are found is rather remarkable.  The plots of the individual parts in Fig. \ref{fig:ising_parts} have behavior where there is a low entropy region, an area of increase, sometimes a small peak, and then a leveling off to a near-maximally entangled plateau.  
The additions and subtractions in Eqs. (\ref{eq:weak}) \& (\ref{eq:strong}) cause the low-entropy regions of the individual bipartite graphs to all balance out in the combination.  Likewise, the high-entropy regions of such graphs all cancel out to low values like the low-entropy regions.  The only interesting feature that remains in combination is a remaining peak, small in comparison to the original plateaus, resulting from the intermittent regions of the bipartite graphs.

\subsection{Lattice $\phi^4$ with Qutrits}
\label{subsec:qutrit}

We next look at the lattice $\phi^4$ model, 
\begin{equation}
    \label{eq:phi4_ham}
    \hat{H}_{anharm}=\sum_{ j} \omega(\hat{a}_{ j}^\dagger \hat{a}_{ j} +\frac{\mathbf{1}}{2})+ \lambda \hat{\phi} _{ j}^4 -2\kappa \hat{\phi}_{ j} \hat{\phi}_{ j+1}.  
\end{equation}
This model is mostly similar to what would be seen in an introductory field theory course where $\lambda$ is the anharmonicity, $\kappa$ the coupling constant, and $\phi$ defined in the usual way.  The difference comes when instituting a digitization of the fields \cite{maxton_digphi}.  We want a highest and lowest energy level such that
\begin{equation}
    \label{eq:restrict_ladder}
    \hat{a} \ket{0}=0 \text{ and } \hat{a}^\dagger \ket{n_{max}-1}=0,  
\end{equation}
with normal operator behavior otherwise.  This is accomplished by a projector in the highest energy state, 
\begin{equation}
    \label{eq:projector}
    P_{n_{max}-1}=\ket{n_{max}-1}\bra{n_{max}-1},  
\end{equation}
which will modify the commutator of the creation and annihilation operators to
\begin{equation}
    \label{eq:proj_comm}
    [\hat{a}, \hat{a}^\dagger]=\mathbf{1}-n_{max}P_{n_{max}-1}.
\end{equation}

\begin{figure}[h!]
\centering
\includegraphics[width=7cm]{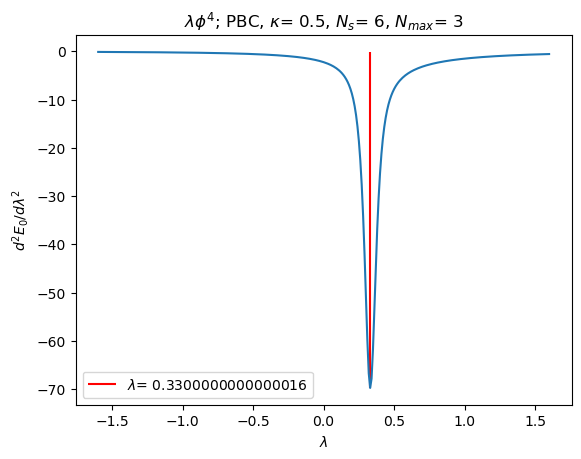}
\caption{\label{fig:phi4_phasepoint}
Identifying the $\lambda$ value where a second-order phase transition would occur for a six-qutrit system.}
\end{figure}

By applying this truncation, we can treat the system as qudits, where $d=n_{max}$.  We specifically restrict ourselves to three energy levels, so it is like a system of qutrits.  This model also exhibits a disordered and ordered phase and can have its basis changed.  We can identify the critical point by taking the second derivative of the ground-state energy with respect to a critical parameter and finding a discontinuity, as shown in Fig. \ref{fig:phi4_phasepoint}.  So, for a six-qutrit system with the given parameters, the critical point should be at $\lambda=.33$.  

\begin{figure}[h!]
\centering
\includegraphics[width=4.25cm]{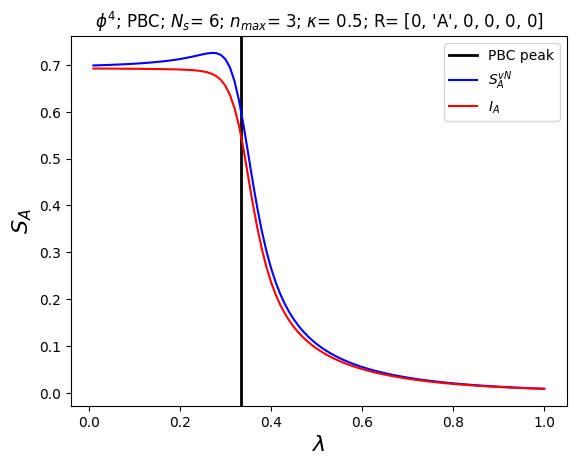} 
\includegraphics[width=4.25cm]{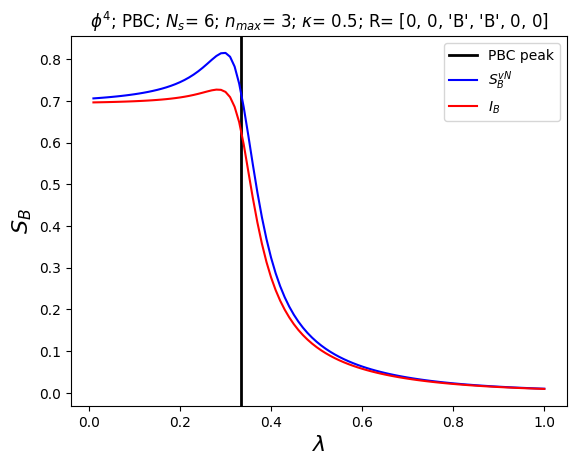} 
\includegraphics[width=4.25cm]{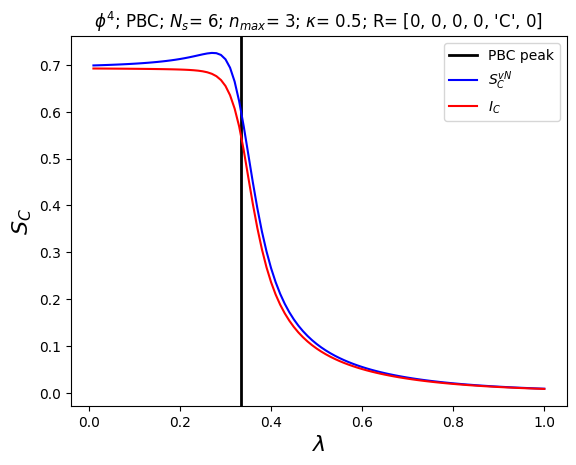} 
\includegraphics[width=4.25cm]{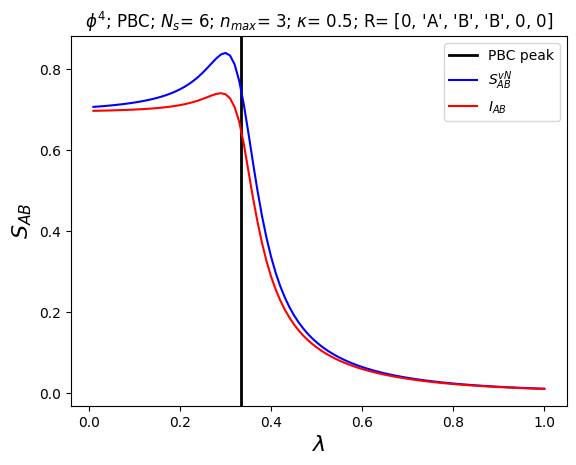} 
\includegraphics[width=4.25cm]{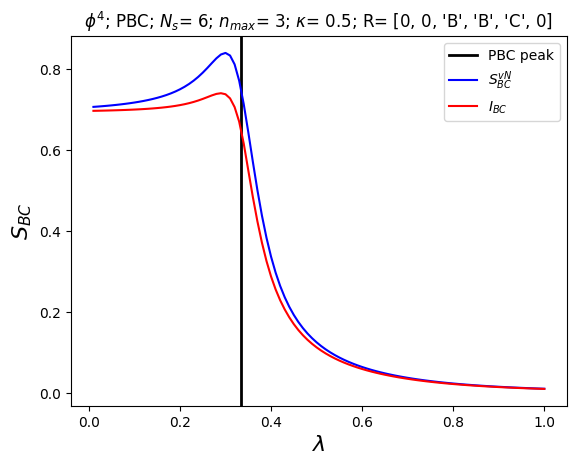} 
\includegraphics[width=4.25cm]{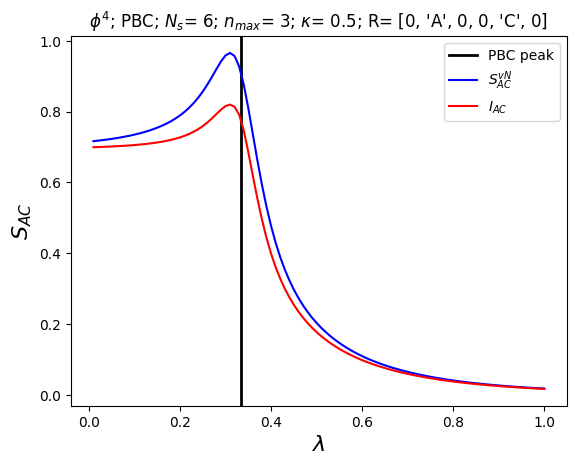}  
\includegraphics[width=4.25cm]{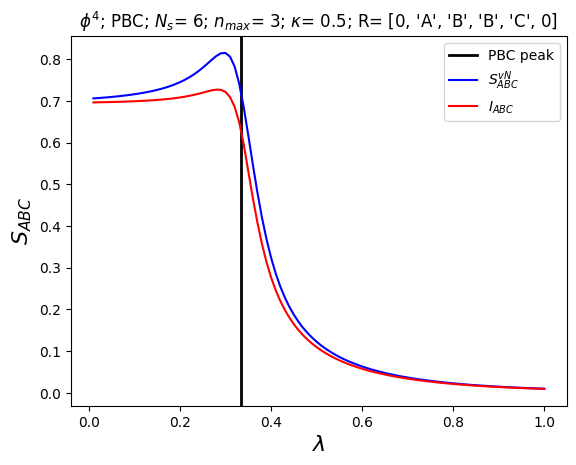}   
\caption{\label{fig:qutrit_parts}6-site qutrit chain varying over $\lambda$ showing multipartite entropy with partitioning DABBCD.  Critical point is marked with vertical black line.  From top left to bottom: $S_A$, $S_B$, $S_C$, $S_{AB}$, $S_{BC}$, $S_{AC}$, $S_{ABC}$.}
\end{figure}

Looking at Fig. \ref{fig:qutrit_parts}, we see very similar behavior to what was seen in Fig. \ref{fig:ising_parts}.  We see that the mutual information captures the behavior in the plots of the entanglement entropy very well.  Again, the places where some deviation occurs are in the sections closer to the critical point in the parameter.  However, now the peaking is not as close to that exact critical point as with some plots for the Ising model.  

We can again show how multipartite entanglement can be used to identify features of the phase change that were not identifiable from just the bipartite quantities.  
\begin{figure}[hbt!]
\centering
\includegraphics[width=4.25cm]{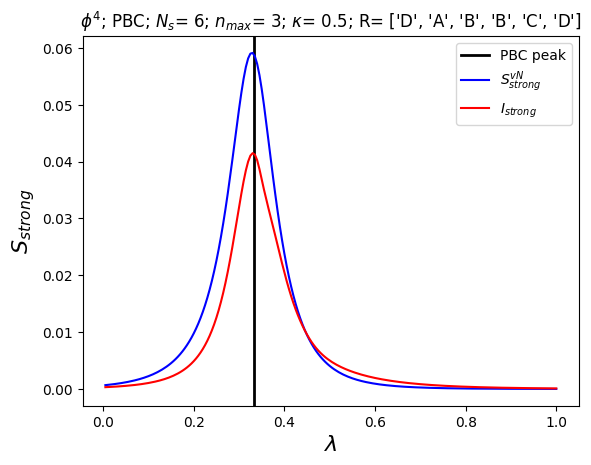} 
\includegraphics[width=4.25cm]{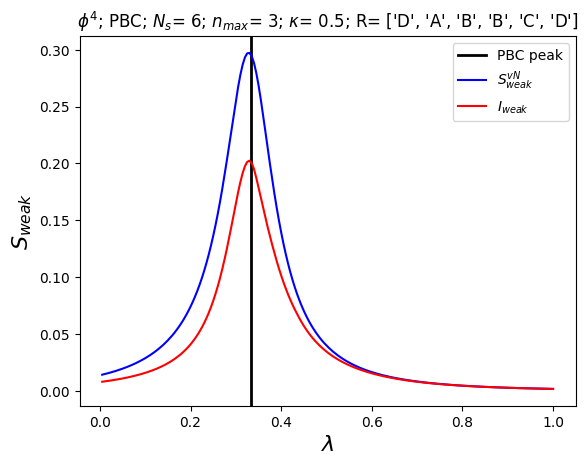}

\caption{\label{fig:qutrit_ineqs}6-site qutrit chain varying over $\lambda$ showing, from left to right, strong subadditivity and weak monotonicity with partitioning DABBCD.}
\end{figure}
The features in Fig. \ref{fig:qutrit_ineqs} retain the same behaviors that were seen for the Ising case.  The mutual information captures the same features of the entanglement entropy with the same overshooting properties.  

\begin{figure}[hbt!]
\centering
\includegraphics[width=7cm]{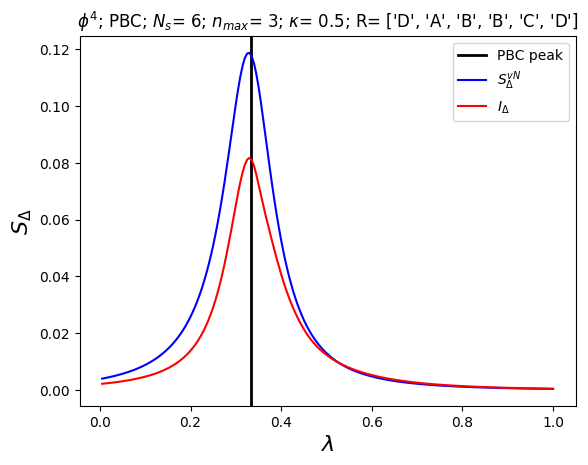}
\caption{\label{fig:qutrit_delta}
$S_{\Delta}$ for a 6-site qutrit chain varying over $\lambda$ with partitioning DABBCD.}
\end{figure}

These quantities combine to give $S_\Delta$, as seen in Fig. \ref{fig:qutrit_delta}.  We observe that the quantity $S_\Delta$ can identify the critical point for the second-order phase transition at the same point as the second derivative method.  Moreover, the mutual information again peaks at a position near the spot of the entanglement, continuing to show promise as an alternative.    

\subsection{Chain of Rydberg Atoms}
\label{subsec:rydchain}
The last model with which we test our methods highlights the current interest in neutral-atom simulation in recent works, the Rydberg chain \cite{rydphase_chain, rydphase_nature}.   Such a system has the Hamiltonian 
\begin{align}
    \label{eq:ryd_ham}
    \hat{H} &= \frac{\Omega}{2}\sum_i(\ket{g_i}\bra{r_i} + \ket{r_i}\bra{g_i}) \notag \\
    &-\Delta\sum_i  \hat{n}_i +\sum_{i<j}V_{ij}\hat{n}_i \hat{n}_j.
\end{align}

Notationally, the system has a ground, $\ket{g}$, and Rydberg, $\ket{r}$, which are analogous to $\ket{0}$ and $\ket{1}$ in gate-based quantum computing.  In the Hamiltonian, $\Omega$ is the Rabi frequency, $\Delta$ is the detuning frequency, $\hat{n}_i=\ket{r_i}\bra{r_i}$ is the Rydberg density operator, and $V_{ij}=\Omega R_b^6/r_{ij}^6$ the van der Waals interaction between atoms $i$ and $j$ a distance $r_{ij}$ apart.  $R_b$ is the Rydberg blockade radius, where the interactions are larger than the Rabi frequency; this makes it unfavorable for more than one atom to be in the Rydberg excited state within this distance.  Unless otherwise stated, $\Omega = 2.5\times2\pi$ MHz, $R_b = 8.375$ $\mu$m, and $\Delta = 3.5\times \Omega = 17.5 \pi$ MHz.   

There are two potential approaches to investigating the phase structure of the Rydberg chain.  We can either hold a value of $R_b/a$ constant and vary $\Delta/\Omega$ which is very similar to the disordered-ordered transitions we have already seen, or we can fix $\Delta/\Omega$ and vary $R_b/a$ and we will cut through several $\mathbb{Z}_p$ phases where the $p$ indicates a modulo $p$ repetition of atoms in the Rydberg state (i.e. $\mathbb{Z}_2$ has an atom in the Rydberg state every other atom).  We will investigate the case where we vary $\Delta/\Omega$ to maintain consistency with the previous examples.  

It is also worth noting that we will be looking at these Rydberg systems while imposing periodic boundary conditions.  We impose these conditions on a linear structure by extending the atoms interactions such that the distances between can be found along the line.  Modeling in this way allows us to retain consistency with the other models.  However, it is more feasible for device implementation to have atoms placed in a ring, which will change the distances.  As a result, periodic implementation would look different on a device and will be discussed in Appendix \ref{sec:app_expryd}.  These differences are marginally different from the implementation shown here and do not affect our procedure, but are still shown in full detail there.  

\begin{figure}[hbt!]
\centering
\includegraphics[width=4.25cm]{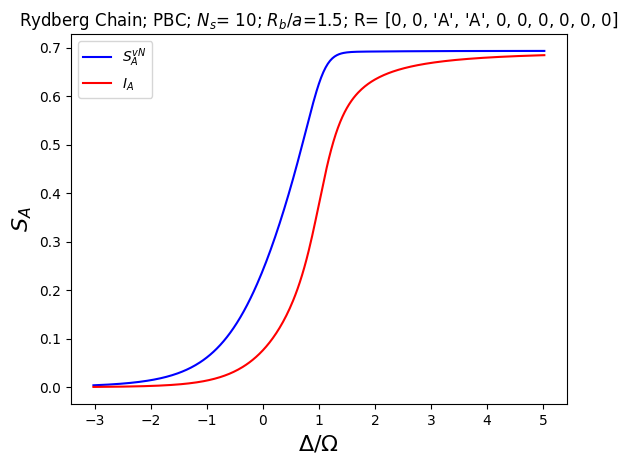} 
\includegraphics[width=4.25cm]{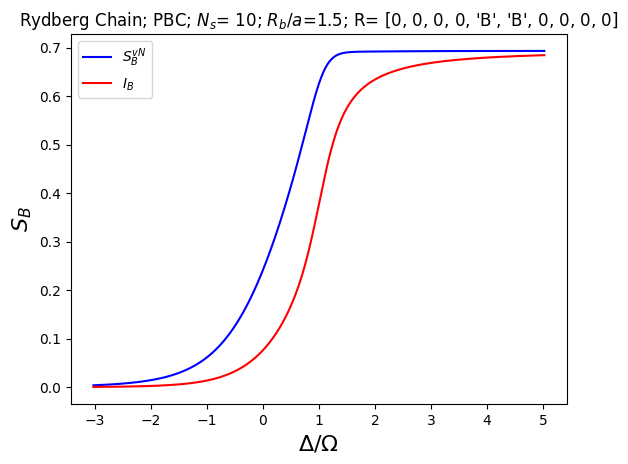} 
\includegraphics[width=4.25cm]{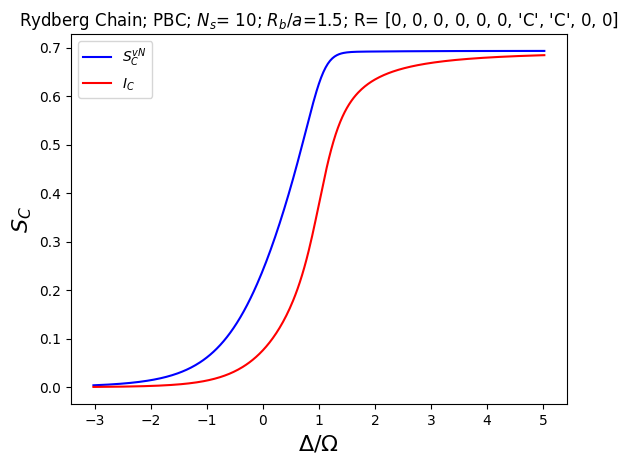} 
\includegraphics[width=4.25cm]{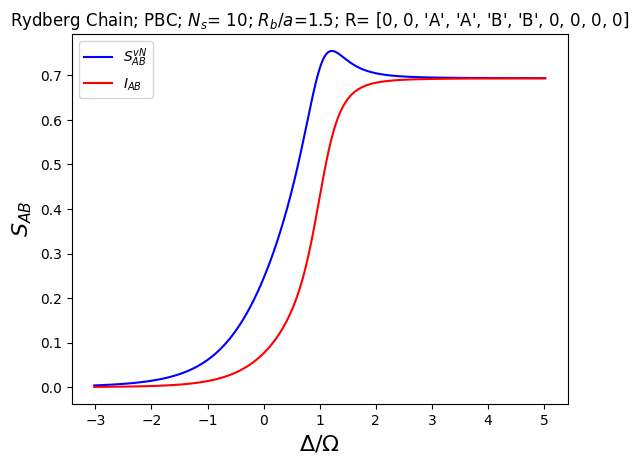} 
\includegraphics[width=4.25cm]{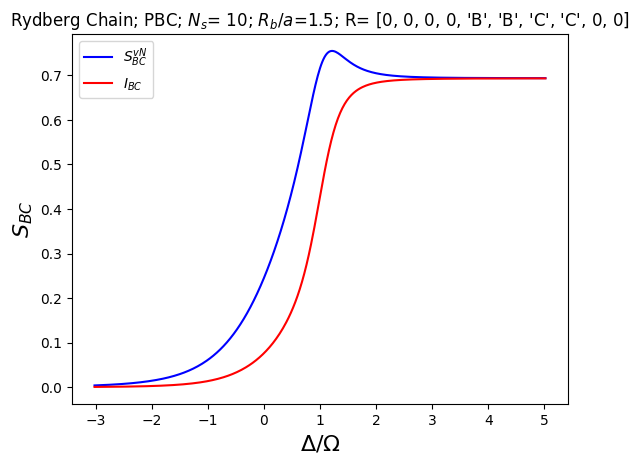} 
\includegraphics[width=4.25cm]{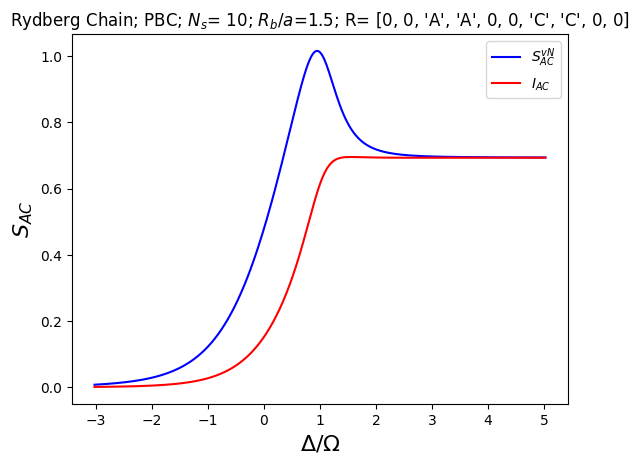} 
\includegraphics[width=4.25cm]{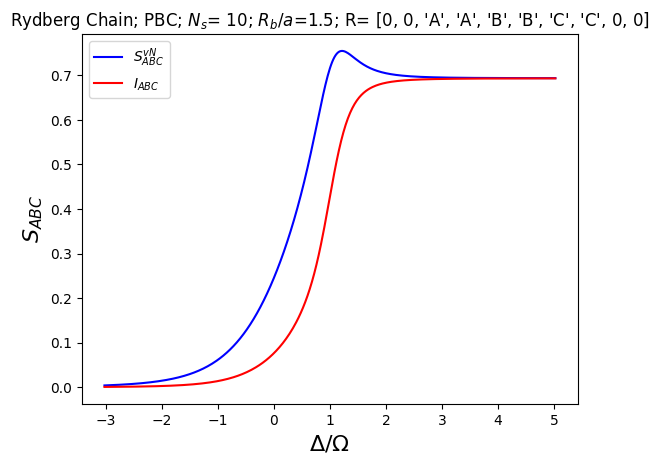}   
\caption{\label{fig:rydchain_parts}10-site Rydberg chain varying over $\Delta/\Omega$, with $R_b/a=1.5$, showing multipartite entropy with partitioning DDAABBCCDD.  From top left to bottom: $S_A$, $S_B$, $S_C$, $S_{AB}$, $S_{BC}$, $S_{AC}$, $S_{ABC}$.
} 
\end{figure}

We are able to again calculate the multipartite entanglement entropy for the system, and approximate it with the mutual information in Fig. \ref{fig:rydchain_parts}.  

We can see the behavior we have come to associate with a disordered-to-ordered phase transition.  An area of low entropy, followed by an area of sharp increase, and then a plateau at higher entropy.  As in the other examples, the mutual information provides a lower bound, and does not easily capture the features of the entanglement spike before leveling off at the higher plateau.  This behavior is much worse in this case for the Rydberg chain compared to previous cases \textemdash we only see the slightest peaking in the $S_{AC}$ case for the mutual information, which exhibits the highest entanglement spike, but nothing noticeable for the other parts.  

We can also test for critical points just as we did with the previous models.  

\begin{figure}[hbt!]
\centering
\includegraphics[width=4.25cm]{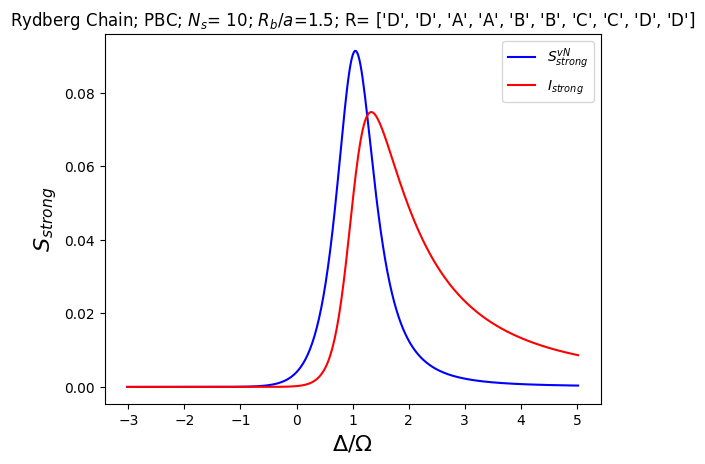} 
\includegraphics[width=4.25cm]{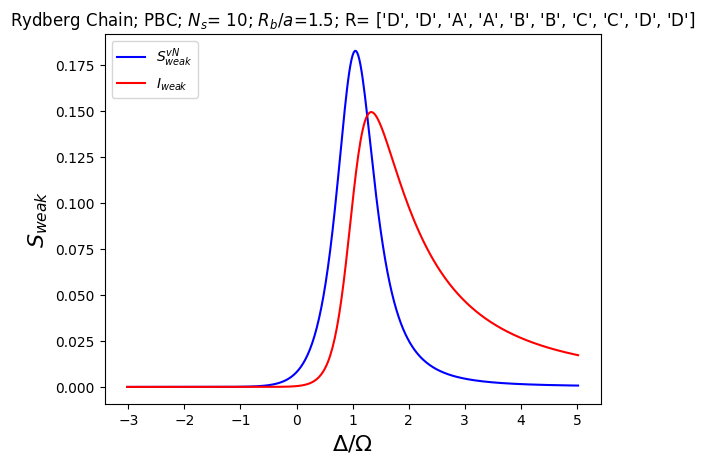} 

\caption{\label{fig:rydchain_ineqs}10-site Rydberg chain varying over $\Delta/\Omega$, with $R_b/a=1.5$, from left to right, strong subadditivity and weak monotonicity with partitioning DDAABBCCDD.}
\end{figure}

\begin{figure}[hbt!]
\centering
\includegraphics[width=7cm]{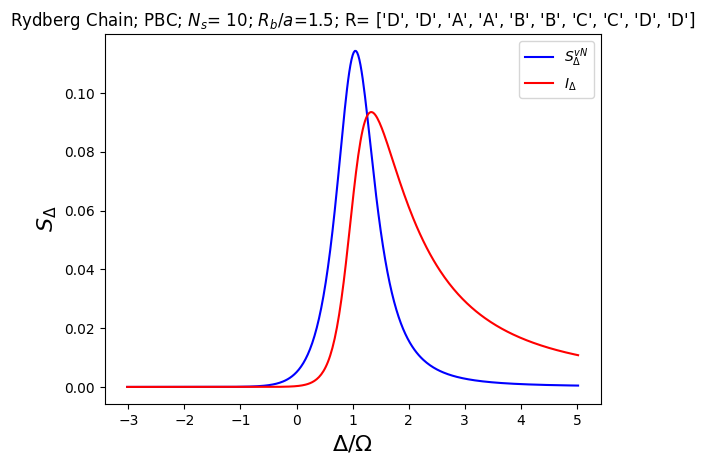} 
\caption{\label{fig:rydchain_delta}
$S_{\Delta}$ for a 10-site Rydberg chain varying over $\Delta/\Omega$, with $R_b/a=1.5$ , and partitioning DDAABBCCDD.}
\end{figure}

Looking at Fig. \ref{fig:rydchain_ineqs}, we observe that the mutual information does not form a tight lower bound of the entanglement peak.  In fact, it is not even a slight difference as in previous cases, it now seems to be peaking at a different place entirely and not having the sharp decline of a peak, like the von Neumann quantity.  Of course, this becomes even more obvious in the composite $S_\Delta$ in Fig. \ref{fig:rydchain_delta}.  

The shift of the mutual information peak compared to the entanglement entropy is natural in the light of the partitioned plots, Fig. \ref{fig:rydchain_parts}, where now the mutual information is no longer as consistently tight of a bound.  The $S_\Delta$, along with the inequalities, work by canceling regions of similar entropy values:  the low-entropy regions stay near zero, all the high plateaus cancel, thus leaving the in-between.  Comparatively, it looks as if the mutual information just does not behave exactly the same as the entanglement in this case, despite still being fairly similar.  

\section{Discussion on Effects of Open Boundary Conditions}
\label{sec:obc}
Up until now we have exclusively looked at entropic calculations with periodic boundary conditions. 
 It is perfectly legitimate to still do these calculations with open boundary conditions.  The difference now being that we are losing the translational invariance that the previous conditions offered.  This will result in some minor shifts of the individual bipartitions' graph, but where we especially see changes is in the quantities of Eqs.  (\ref{eq:sdelt}), (\ref{eq:weak}), \& (\ref{eq:strong}) where the loss of this invariance greatly affects the overall quantity across the varied parameter.  

\begin{figure}
     \centering
     \includegraphics[width=4.25cm]{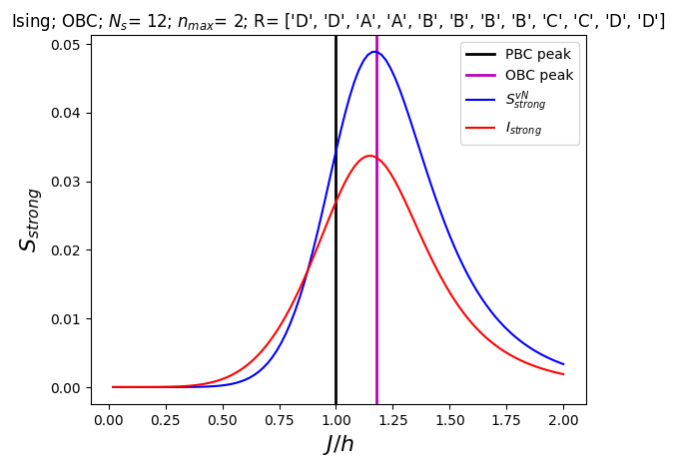}
     \includegraphics[width=4.25cm]{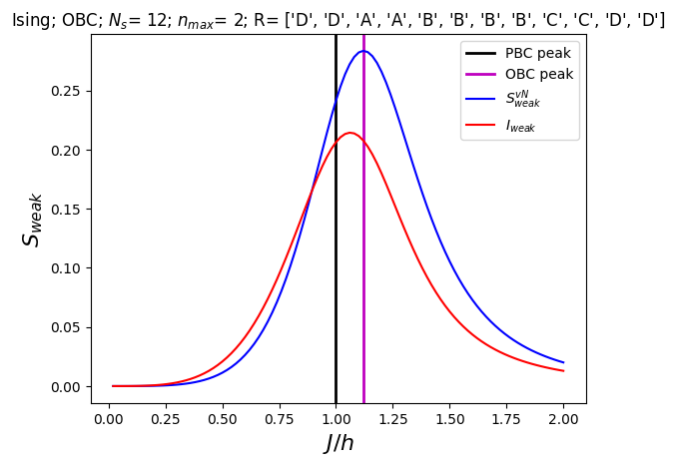}
     \includegraphics[width=4.25cm]{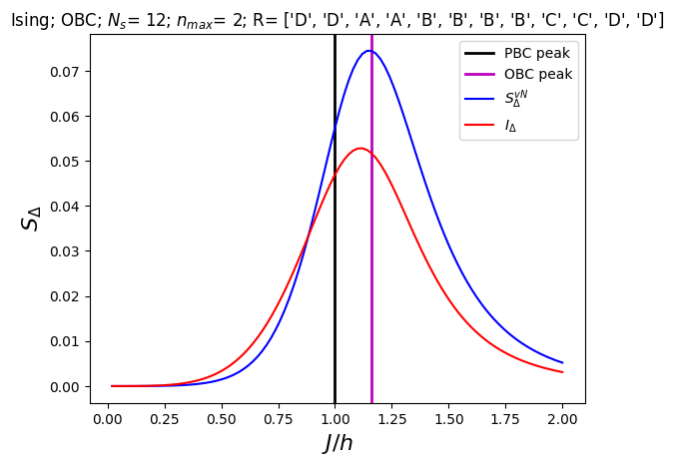}
     \caption{From left to right, for Ising, strong subadditivity, weak monotonicity, and $S_\Delta$ over $J/h$.}
     \label{fig:ising_obc}
\end{figure}
\begin{figure}
     \centering
     \includegraphics[width=4.25cm]{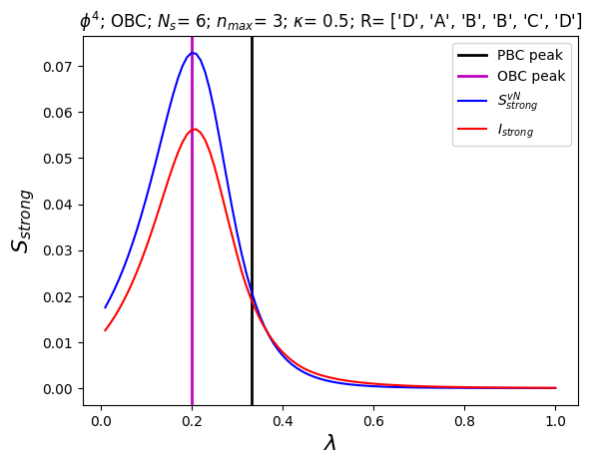}
     \includegraphics[width=4.25cm]{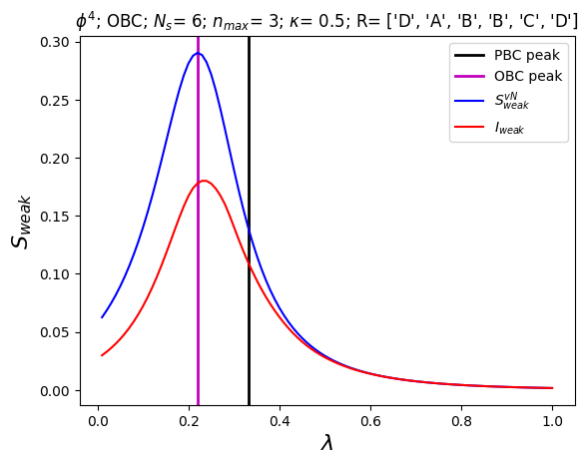}
     \includegraphics[width=4.25cm]{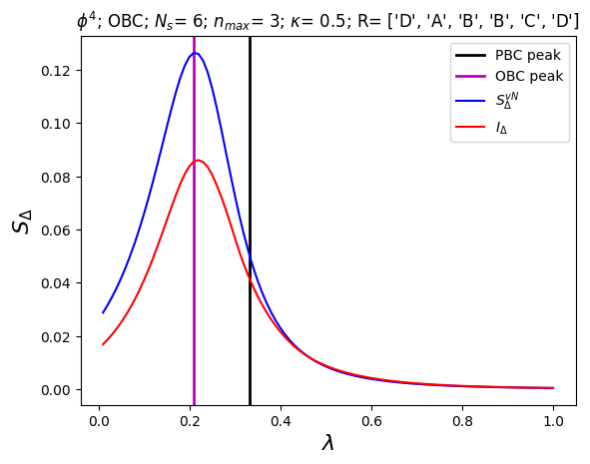}
     \caption{From left to right, for qutrit $\phi^4$, strong subadditivity, weak monotonicity, and $S_\Delta$ over $\lambda$.}
     \label{fig:qutrit_obc}
\end{figure}

In Figures \ref{fig:ising_obc} \& \ref{fig:qutrit_obc}, we see that the major effect of the open boundary conditions is an overall shift in the position of the quantity's peak towards the disordered phase of the periodic boundary condition graph.  As both $S_{strong}$ and $S_{weak}$ are composed of different bipartite entropy quantities, it follows naturally that these different bipartite entropies are also shifted as a result of the new boundary conditions.  

To look closer at the difference in boundary conditions for a Rydberg chain, it is beneficial to look at a heatmap of the phases across both parameter ranges.  

\begin{figure}
    \centering
    \includegraphics[width=7.cm]{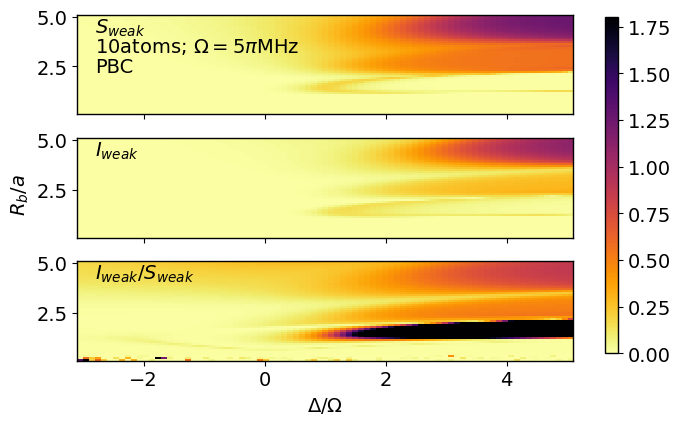}
    \includegraphics[width=7.cm]{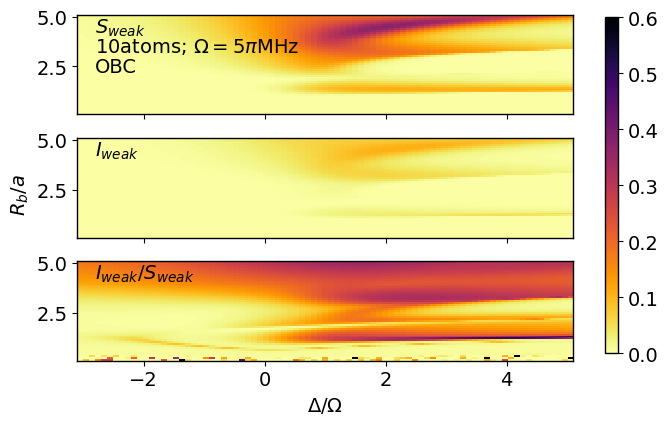}
    \caption{Phase maps looking at the weak monotonicty for a Rydberg chain system of partitioning DDAABBCCDD.  Comparing the entanglement, mutual information, and their ratio between open and periodic boundary conditions.  }
    \label{fig:rydchain_conds}
\end{figure}

Looking at Figure \ref{fig:rydchain_conds} highlights how much the boundary conditions impact the overall phase structure of the model.  In the periodic case, there are three well-defined lobes.  However, when looking at the open boundary conditions, we see that the cleanliness of the structure is lost.  There is also a change in where the high-entropy regions are, along with an overall decrease in the scale of entropy.  This makes apparent the importance of boundary conditions in the calculation of composite entanglement quantities.  This follows naturally from the fact that for periodic boundary conditions any entanglement between sites on opposite ends of a configuration is nearest-neighbor, but in open boundary conditions, they are now on completely different sides of the system.  

Further discussions surrounding how open boundary conditions effect the calculation of critical points using $S_{\Delta}$ are discussed in Appendices \ref{sec:app_epsising} \& \ref{sec:app_sdelt_dep}.

\section{Filtered Probabilities for Previous Models}
\label{sec:filt}
Despite being a strict lower bound on the von Neumann entanglement entropy, it is clear that the mutual information still has room for improvement towards reaching the exact value.  In an effort to use ditstring probabilities to get even closer to the entanglement, a method of filtering out low-probability states has been proposed as a way of improving the bound \cite{filter_bits}.  This scheme works by choosing a minimum probability value and then throwing out all probabilities lower than it, then renormalizing the remaining probabilities and going on to calculate the Shannon entropies from there.  However, in this aforementioned work, the method was only applied to bipartitions of Rydberg ladder systems.  

\begin{figure}[htp]
    \centering
    \includegraphics[width=4.25cm]{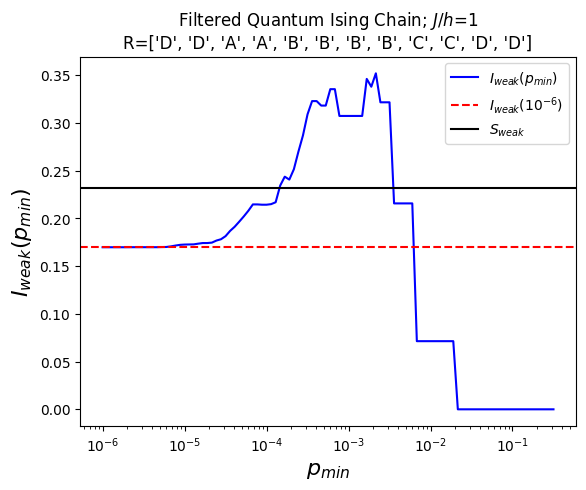} \hfill
    \includegraphics[width=4.25cm]{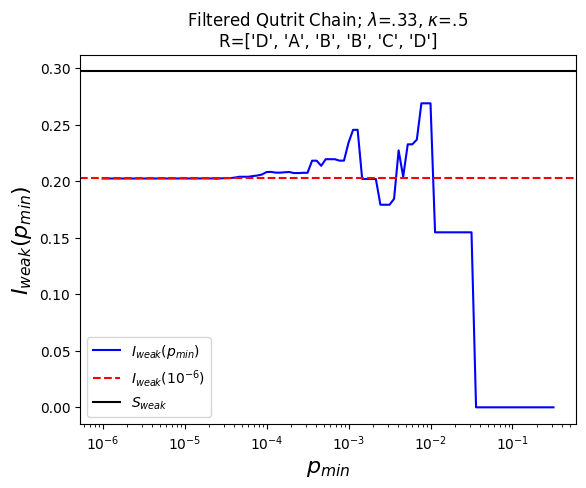}\hfill
    \includegraphics[width=4.25cm]{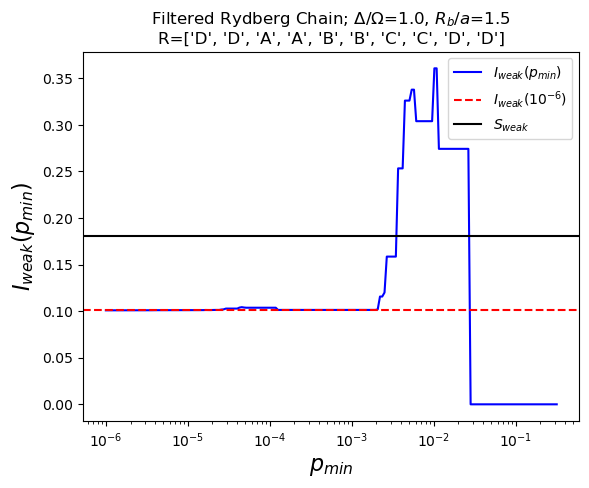}
    \caption{From left to right, filtering applied to Ising, $\phi^4$, and Rydberg chain systems.}
    \label{fig:filter}
\end{figure}

We sought to apply this truncation scheme to different models and a composite quantity, where we arbitrarily chose to look at the weak monotonicity. We went into the test with the assumption that due to the additions and subtractions of multiple filtered bipartite entropies, the behavior may be unseemly compared to the results in \cite{filter_bits}.  

However, looking at Fig. \ref{fig:filter} we can see that the behavior is actually encouraging.  All plots exhibit the behavior of an increase in a filtered mutual information above the unfiltered mutual information.  These graphs also exhibit plateaus at certain mutual information values across probability ranges, which is typical.  In each graph there is an area of increase in mutual information with filtering, then an area where the quantity dips, then rises again, all before the final drop to zero as all probabilities are filtered out.  Typically, such graphs just exhibit an area of increase and then a decrease to zero.  There are exceptions to the rule, but not in as distinct ways as these.  We believe that this could be due to the additions and subtractions of multiple quantities.   

Currently, work is underway to identify an optimal filtering level for these composite quantities using inflection points of their conditional entropy, as detailed in \cite{filter_bits}.  Early returns are promising and will be discussed in a future work.  The positive response of $S_{weak}$ to filtration is indicative of the feasibility of utilizing our method of finding critical points with peaking quantities combined with the mutual information.  A filtration level is akin to only having ditstring results of a quantum computer matching that total counts.  For example, a filtration level of $10^{-3}$ for exact diagonalization presents an idealized return for a 1000 shot trial on a quantum computer.  This kind of idea is further tested and expanded upon for Rydberg ladder systems in \cite{exploring}.  As another attempt to improve the mutual information towards the exact entanglement value, the change of computational basis is investigated in Appendix \ref{sec:app_mutinf}.    

\section{Conclusions}
\label{sec:conc}

We have demonstrated that the mutual information, despite involving the combination of multiple individual bipartite quantities, still operates as a lower bound to the entanglement entropy and still suitably captures the behavior of the entanglement entropy for multipartite quantities.  We have also shown that the composite multipartite quantities of \cite{sdelta} can suitably identify critical points with their extrema across a varied parameter, by canceling low- and high-entropy regions to either side of this transition point in the individual entropy calculations that compose the quantity.  We push the mutual information approximation further, by using it for these multipartite entanglement quantities, and find that it performs reasonably well and peaks in near the same spot.  Further work must be completed to better elucidate the root of this behavior and to determine if there are limits that can be taken to make the mutual information exactly in line.  Regardless, the mutual information still approximates the behavior of the maximum in the entanglement for this purpose.  We have also shown the possibilities for finding such multipartite quantities with quantum computers by seeing the viable responses of the multipartite composite quantities to filtering methods.  

Work is in progress for extending these methods to tests beyond 1+1D, specifically work towards an extension in square arrays of Rydberg atoms.  We will also need to better understand the underlying behavior that causes the peaks of $S_{weak}$, $S_{strong}$, and $S_\Delta$ to be at the locations of critical points.  

\section*{Acknowledgements} This research was supported in part by the Dept. of Energy
under Award Number DE-SC0019139 and DE-SC0010113. 
We thank A. Maloney, S. Catterall, M. Asaduzzaman, B. Senseman, A. Kaufman, G. Can Toga, A. Samlodia, J. McGreevy, and Ting-Chung Lin for discussions. 
This work benefited from the Co-design for Fundamental Physics in the Fault-Tolerant Era  IQus workshop (April 2025), which was supported by U.S. Department of Energy, Office of Science, Office of Nuclear Physics, InQubator for Quantum Simulation (IQuS) under Award Number DOE (NP) Award DE-SC0020970 via the program on Quantum Horizons: QIS Research and Innovation for Nuclear Science.

\section*{Data Availability}  The datasets generated for all figures were done with exact diagonalization and are available in the Zenodo repository \cite{zenodo}.

%

\clearpage
\appendix
\section{Basis Dependence of the Mutual Information}
\label{sec:app_mutinf}
\subsection{1D Quantum Ising}

The mutual information is a lower bound, but leaves room for improvement.  We have two representations for the 1D Quantum Ising model and can go between them with a basis change.  One could guess that one representation may be better than the other, or mirroring known behavior, maybe the X basis is better in the ordered phase than the Z.  By rotating between the bases with 
\begin{equation}
    \label{eq:ising_rot}
    U=e^{\frac{i}{2}\sigma^y \phi},
\end{equation}
we can test this.  

\begin{figure}[hbt!]
\centering
\includegraphics[width=4.25cm]{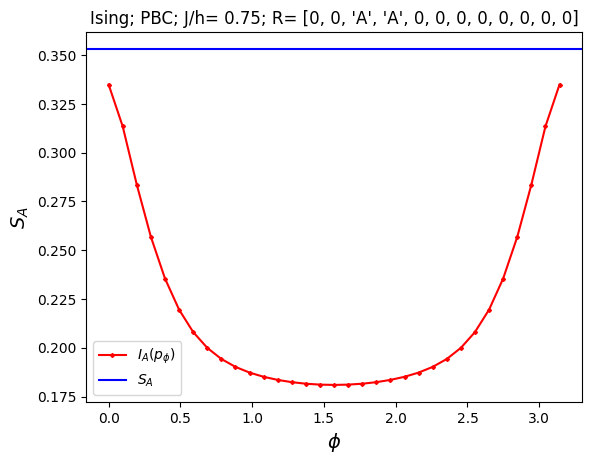} 
\includegraphics[width=4.25cm]{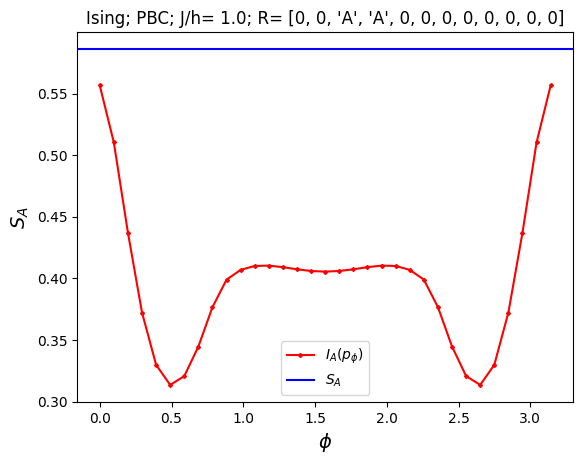}
\includegraphics[width=4.25cm]{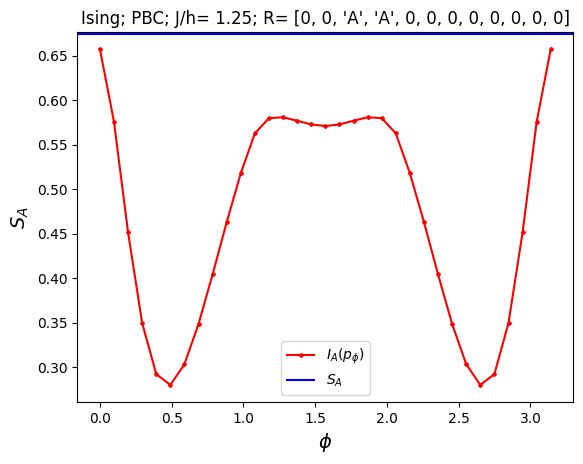}
\caption{\label{fig:ising_basis}For a 12-site Ising chain, looking at an $|A|=2$, $S_A$ for all, rotating the basis through $\pi$ radians.  From left to right with respect to $J/h$, the disordered phase, at criticality, the ordered phase.}
\end{figure}

From Fig. \ref{fig:ising_basis} we see that the Z basis is clearly the best option in the disordered phase, and the X basis continues to improve as $J/h$ increases.  However, it does not ever surpass the Z basis as an approximation for the entanglement entropy, not even in the ordered phase.  
This seems to indicate a potential best basis to be used for the mutual information in comparison to the entanglement.  

\begin{figure}
    \centering
    \includegraphics[width=7.cm]{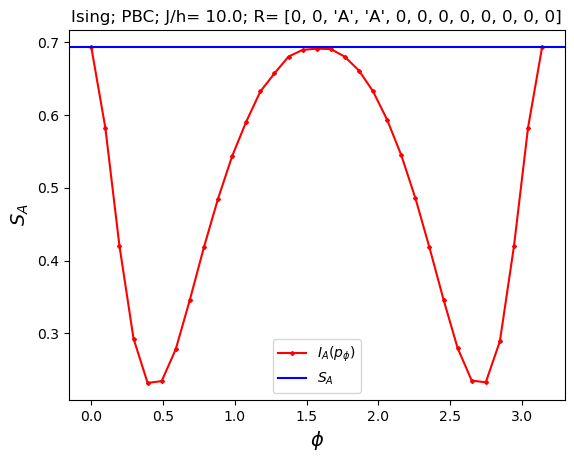}
    \caption{For a 12-site Ising chain, looking at an $|A|=2$, $S_A$. We rotate the basis at a large $J/h$.}
    \label{fig:big_ising_basis}
\end{figure}

While it may seem that if we were to continue increasing $J/h$ we could eventually get an instance where the X basis would be a better basis choice than the Z basis, this is not the case.  In Fig. \ref{fig:big_ising_basis}, we take $J/h$ very large, and we can see that at best the X basis for the mutual information can be said is comparable to the Z basis in capturing the entanglement entropy, but it still does not come as close to approximating the entanglement as the Z basis.  

\subsection{Lattice $\phi^4$ with Qutrits}
\begin{figure}[hbt!]
\centering
\includegraphics[width=4.25cm]{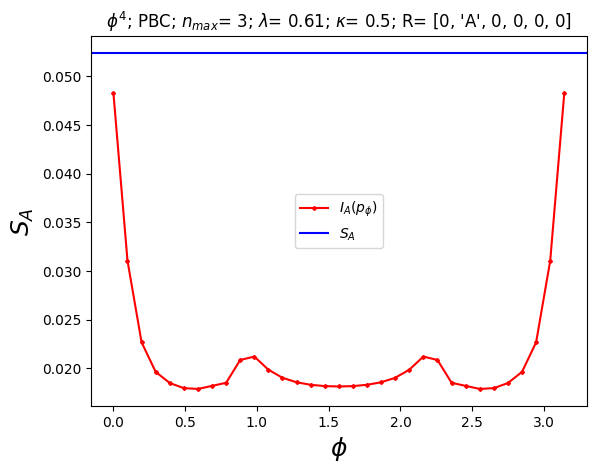} 
\includegraphics[width=4.25cm]{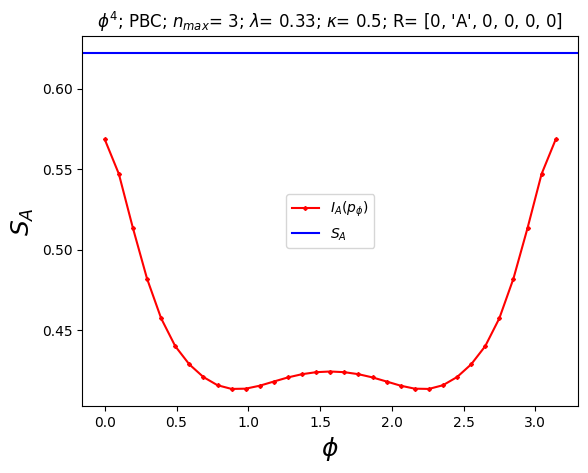}
\includegraphics[width=4.25cm]{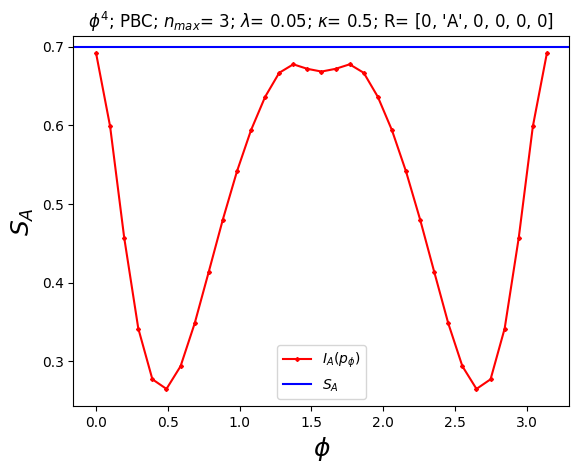} 
\includegraphics[width=4.25cm]{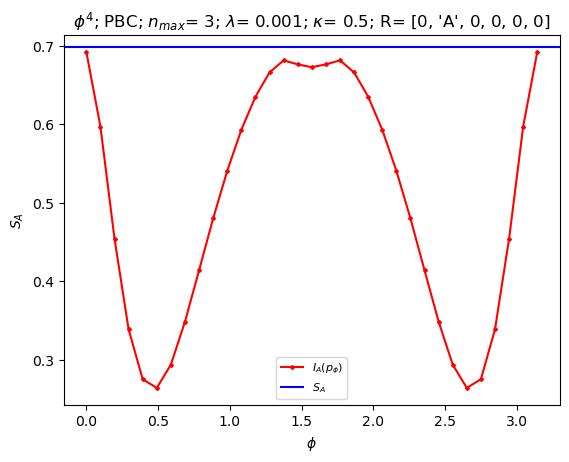}
\caption{\label{fig:qutrit_basis}For a 6-site qutrit chain, looking at an $|A|=1$, $S_A$ for all, rotating the basis through $\pi$ radians.  From left to right with respect to $\lambda$, the disordered phase, at criticality, the ordered phase. The last plot is taken deep in the ordered phase to mirror what was done in Fig. \ref{fig:big_ising_basis}.}
\end{figure}

Again, as with the Ising case, we can look at how a change of basis would impact the mutual information, specifically compared to the von Neumann entanglement as a lower bound for lattice $\phi^4$.  Since our model can approximate qudits, we need to utilize a general spin operator (as opposed to the spin-1/2 operator used for Ising),  
\begin{equation}
    \label{eq:unitary_rot}
    U=e^{iS_y \phi}.  
\end{equation}
We again observe similar behavior to the basis rotation in the Ising model.  Looking at Fig. \ref{fig:qutrit_basis}, we again start in the Z basis and rotate into the X basis, then back to the Z basis.  Again, we see that regardless of which side of criticality the parameter is chosen from, the Z basis is always the best basis to be used for the mutual information.

For consistency with our Ising investigation, we take the small $\lambda$ limit for the qutrit model in the fourth plot of Figure \ref{fig:qutrit_basis}.  Here we again see that our basis-rotated mutual information gets closer to the entanglement entropy at the near $\pi/2$ rotation, but still will not approximate as well as the initial Z basis.  

\section{$S_{\Delta}$ Dependency on Boundary Conditions}
\label{sec:app_epsising}

To highlight how the location of the critical point shifts in the parameter range with boundary conditions, we use what we refer to as epsilon boundary conditions which allows us to interpolate  between open and periodic boundary conditions gradually.  This is implemented with a 1D Quantum Ising Hamiltonian by,

\beq
    \label{eq:ising_eps}
        \hat{H}_{\text{Ising, }\epsilon} = - h \sum_{i=1}^{N_s} \hat{\sigma}^z_i -J \sum_{i=1}^{N_s} \hat{\sigma}^x_{i} \hat{\sigma}^x_{i+1} -J(1-\epsilon)\hat{\sigma}^x_{N_s} \hat{\sigma}^x_{1},
\enq
where when $\epsilon=1$ we are in open boundary conditions and when $\epsilon=0$ we are in periodic boundary conditions, and all other parameters are the same as defined previously.  
\begin{figure}[hbt!]
\centering
\includegraphics[width=7cm]{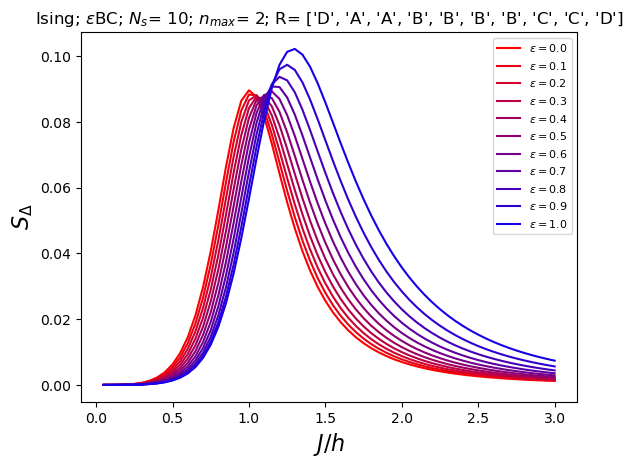} 
\caption{\label{fig:ising_epsilon}
$S_{\Delta}$ for a 10-site Ising chain varying over $J/h$, and partitioning DAABBBBCCD.}
\end{figure}

Looking at Fig. \ref{fig:ising_epsilon}, it is apparent how the change from open to periodic boundary conditions clearly shifts the peak closer to the infinite-volume critical point of $J/h=1$, emphasizing the importance of choice of boundary conditions for the method.  In the next appendix, we will look at more minutiae that impact the peaks identifying critical points.

\section{$S_{\Delta}$ Dependencies}
\label{sec:app_sdelt_dep}

We saw previously that boundary conditions causes a shift in the identification of the critical point.  It was also observed in the main text that although the peaks get incredibly close to the critical point, they do not match it exactly.  In this appendix, we will show for both open and periodic boundary conditions, using the 1D Quantum Ising model, how this method with the entanglement entropy is limited in different ways by finite sizing.  

\begin{figure}[h!]
\centering
\includegraphics[width=4.250cm]{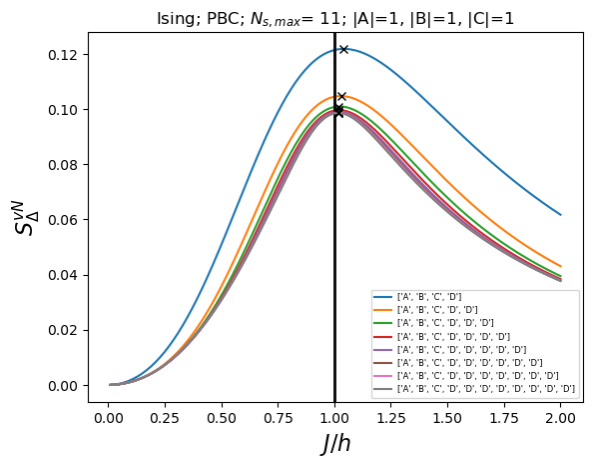} 
\includegraphics[width=4.250cm]{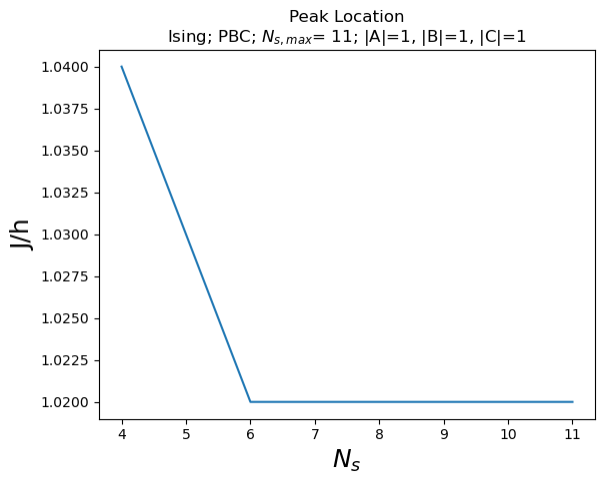}
\includegraphics[width=4.250cm]{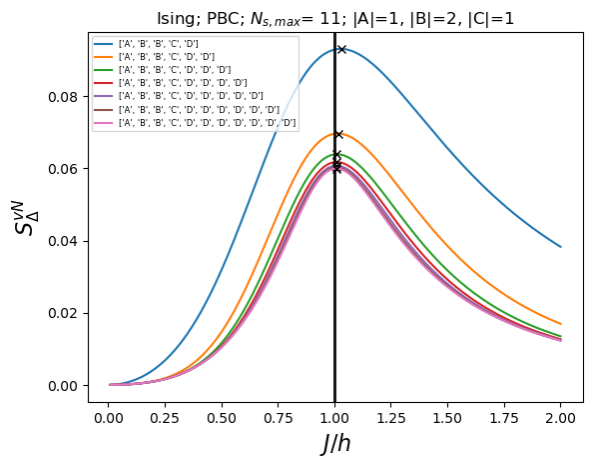}
\includegraphics[width=4.250cm]{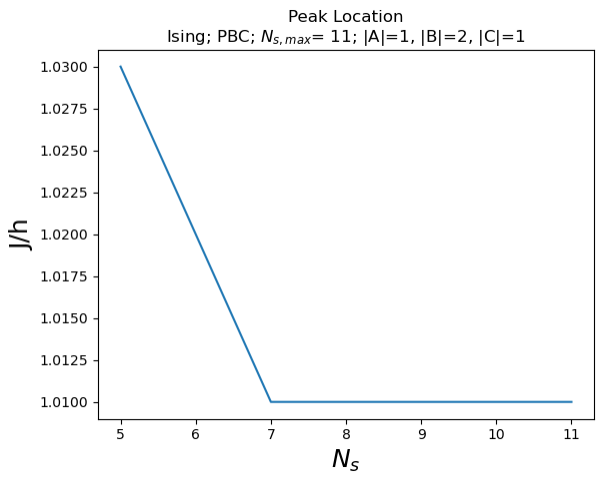}
\includegraphics[width=4.250cm]{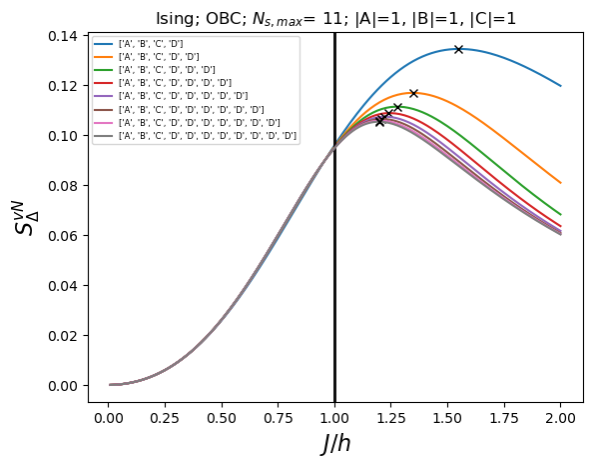}
\includegraphics[width=4.250cm]{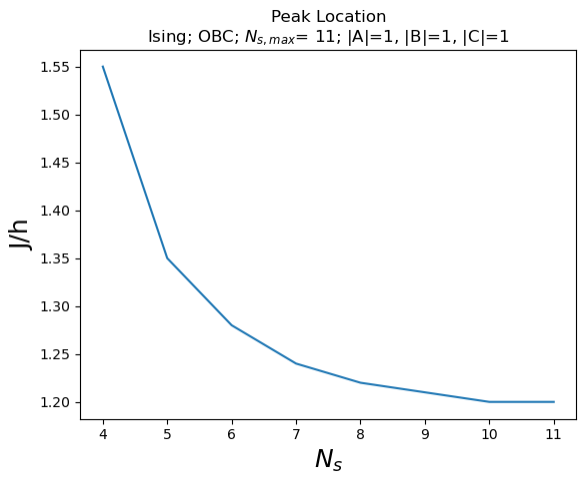}
\includegraphics[width=4.250cm]{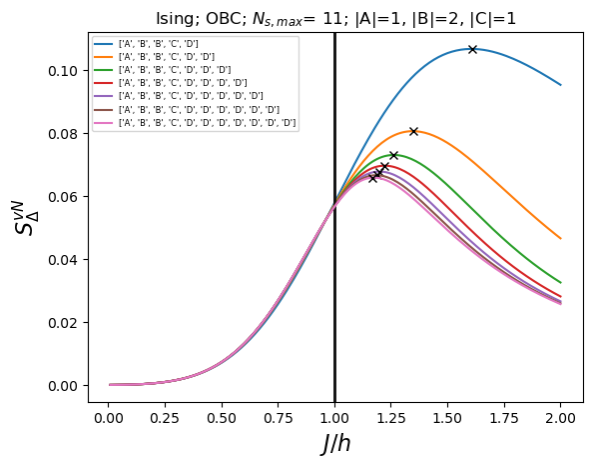} 
\includegraphics[width=4.250cm]{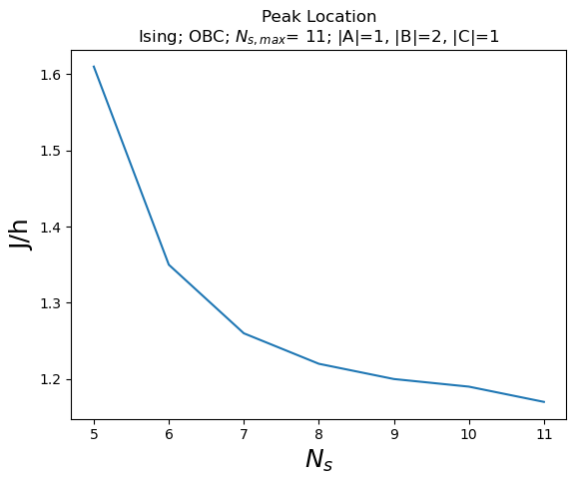}
\caption{\label{fig:ising_sdelt_depend}$S_{\Delta}$ for an Ising chain responding to different system sizes, partition sizes, and boundary conditions.  Location of peaks marked with a black $\times$, and infinite-volume critical point denoted with black vertical line.  }
\end{figure}

In Figure \ref{fig:ising_sdelt_depend}, we look at a few different small changes to see how the critical point adjusts. In all plots, we start with a small system split into $A,B,C,\text{ and }D$ subsystems indicated in the plot titles and $|D|=1$.  Then, the same subsystems remain the same, except $|D|$ incrementally increases by one.  This allows to test both how the critical points respond to increases in overall system sizes and overall subsystem sizes.  Furthermore, these same tests are applied for systems under both periodic and open boundary conditions.  

\begin{figure}[h!]
\centering
\includegraphics[width=4.25cm]{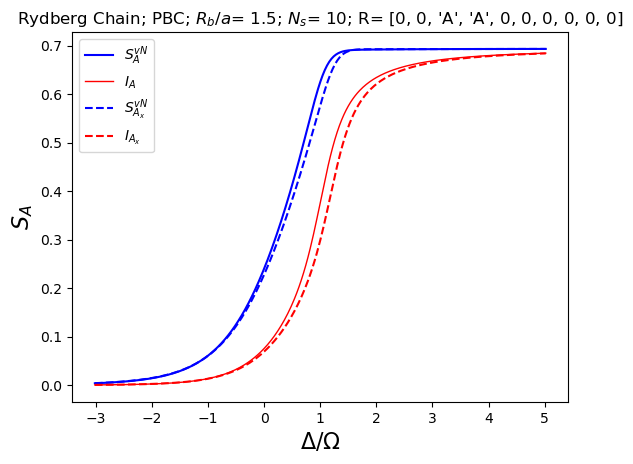} 
\includegraphics[width=4.25cm]{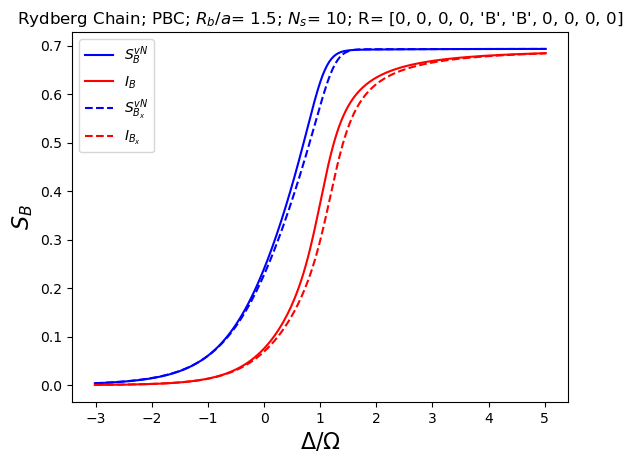} 
\includegraphics[width=4.25cm]{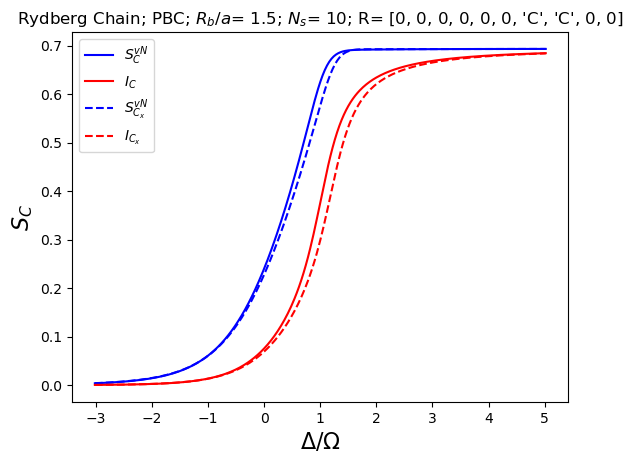} 
\includegraphics[width=4.25cm]{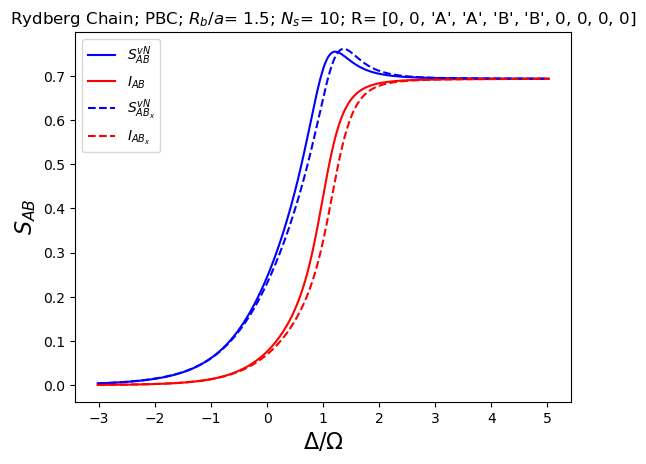} 
\includegraphics[width=4.25cm]{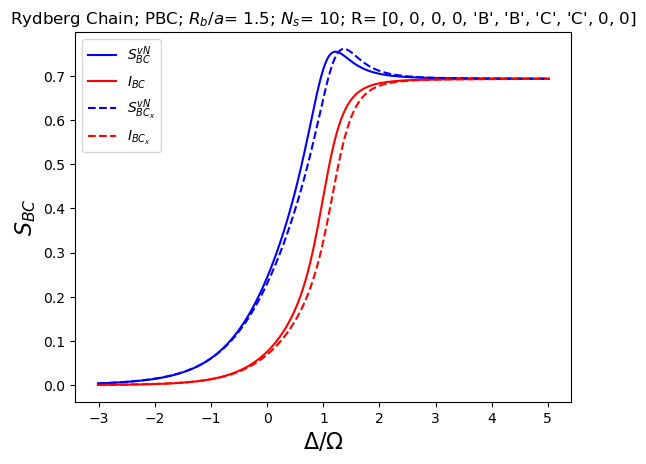} 
\includegraphics[width=4.25cm]{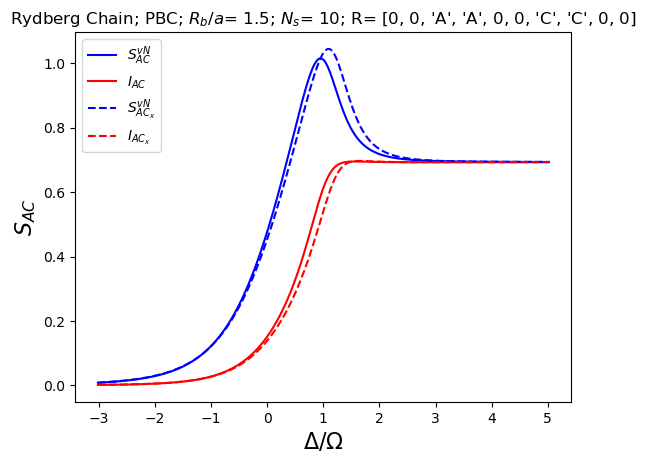} 
\includegraphics[width=4.25cm]{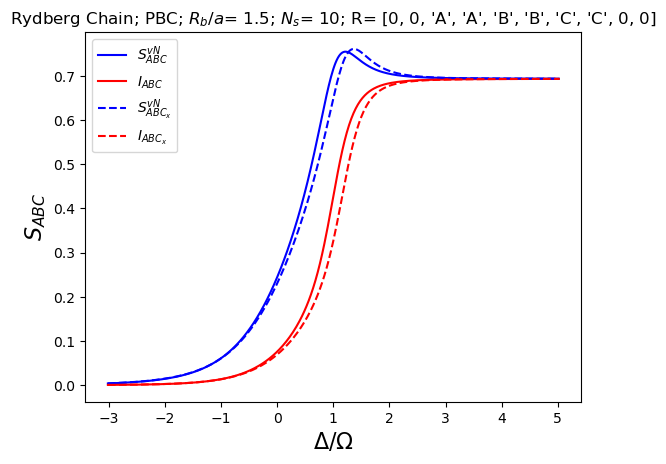}   
\caption{\label{fig:rydchain_exp_parts}10-site Rydberg chain varying over $\Delta/\Omega$, with $R_b/a=1.5$ ,showing multipartite entropy with partitioning DDAABBCCDD.  From top left to bottom: $S_A$, $S_B$, $S_C$, $S_{AB}$, $S_{BC}$, $S_{AC}$, $S_{ABC}$.  Solid lines indicate the periodic boundary conditions procedure outlined in the main text (linear), whereas dashed lines indicate the procedure discussed in this appendix (ring).  
} 
\end{figure}

It is immediately apparent, for both open and periodic boundary conditions, that the peak of $S_{\Delta}$ moves closer to the true critical point with an increase in system size.  This action is more evident in open boundary conditions, however in periodic boundary conditions the peaks are closer to begin with.  It is also clear from tracking peak location in both boundary conditions that by increasing the sizes of subsystems, the overall $S_{\Delta}$ quantity can approach closer to the true critical point than with smaller subsystem sizes.  These series of plots also indicate the limitations of the method in open boundary conditions versus periodic boundary conditions. 

Overall, this informs us that when trying to use $S_{\Delta}$ to identify a system's critical point we want to be using periodic boundary conditions, as large of a system size as possible, and to have subsystem sizes that keep any subsystem from being too large compared to the others.  

\begin{figure}[h]
     \centering
     \includegraphics[width=4.25cm]{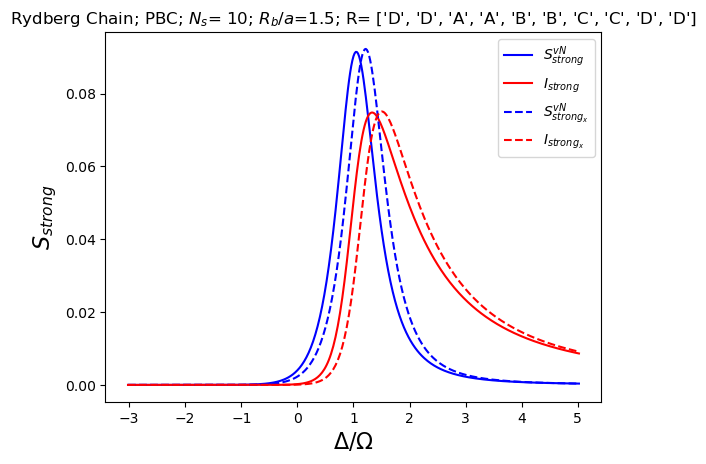}
     \includegraphics[width=4.25cm]{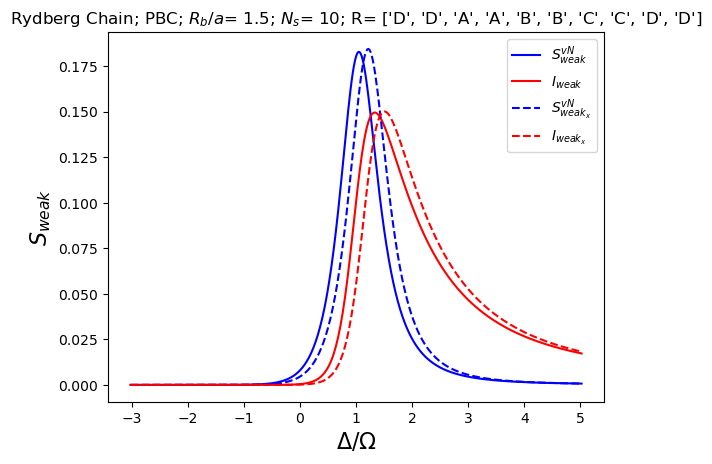}
     \includegraphics[width=4.25cm]{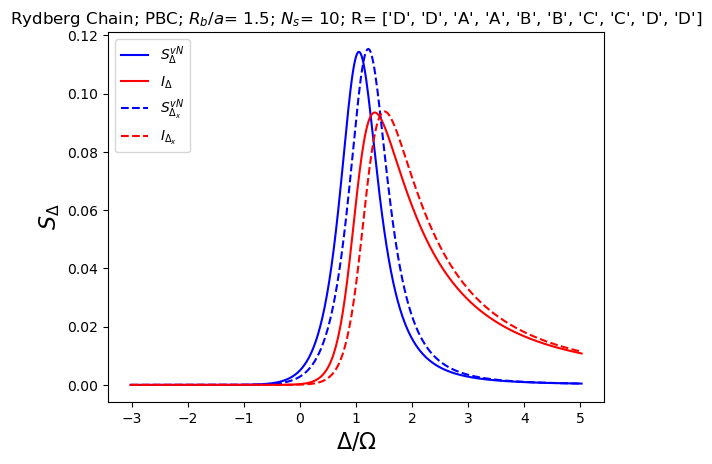}
     \caption{From left to right and top down, for a Rydberg Chain, strong subadditivity, weak monotonicity, and $S_\Delta$ over $\Delta/\Omega$ with $R_b/a=1.5$.  Solid lines indicate the periodic boundary conditions procedure outlined in the main text (linear), whereas dashed lines indicate the procedure discussed in this appendix (ring).}
     \label{fig:rydchain_exp_peaks}
\end{figure}

\section{Rydberg Chain with Device Implementation}
\label{sec:app_expryd}
Previously in the manuscript, we studied a chain of Rydberg atoms under periodic boundary conditions.  We implemented this with straight-line interactions between atoms, to remain consistent with the other chains we studied, as they were limited by nearest-neighbor interactions.  However, on an actual analog quantum device, implementing periodic boundary conditions with Rydberg atoms would not have linear interactions, as this system would be implemented as a closed ring of atoms a fixed distance apart along the circumference.  In this implementation, the interactions would not be taken along the circumference, but rather the shortest distance between two atoms would be chords across the ring.

This adjustment will produce slightly different results from the previous.  This is highlighted in Fig. \ref{fig:rydchain_exp_parts}, where we immediately see small shifts towards higher $\Delta/\Omega$ in both the entanglement entropy and mutual information.  
Naturally, this causes a small shift in the location of the peaks in the composite quantities, as seen in Figure \ref{fig:rydchain_exp_peaks}.  Again,  all the plots of the new implementation of periodic boundary conditions are shifted into a higher $\Delta/\Omega$ range, indicating a different critical point in the ring implementation than is seen with the linearly implemented interactions as seen in \ref{subsec:rydchain}.

\end{document}